\documentstyle[11pt,aaspp4]{article}

\lefthead{Smith et al.}
\righthead{Chemical Compositions in the LMC}

\begin{document}

\title{Chemical Abundances in Twelve Red Giants of the Large Magellanic Cloud 
from High-Resolution Infrared Spectroscopy\footnote{Based on observations
obtained at the Gemini Observatory, which is operated by the Association
of Universities for Research in Astronomy, Inc., under a cooperative
agreement with the NSF on behalf of the Gemini partnership: the National
Science Foundation (United States), the Particle Physics and Astronomy
Research Council (United Kingdom), the National Research Council (Canada),
CONICYT (Chile), the Australian Research Council (Australia),
CNPq (Brazil), and CONICET (Argentina).} }

\author{Verne V. Smith\footnote{Visiting Astronomer, Gemini South
Observatory}}
\affil{Department of Physics, University of Texas El Paso, El Paso, TX 79968 
USA; verne@barium.physics.utep.edu}

\author{Kenneth H. Hinkle}
\affil{National Optical Astronomy Observatory\footnote{Operated by
Association of Universities for Research in Astronomy, Inc., under
cooperative agreement with the National Science Foundation}, 
P.O. Box 26732, Tucson, AZ 85726 USA; khinkle@noao.edu}

\author{Katia Cunha$^{2}$}
\affil{Observatorio Nacional, Rio de Janeiro, Brazil; kcunha@on.br}

\author{Bertrand Plez}
\affil{GRAAL, Universite Montpellier II, Montpellier, France;
plez@graal.univ-montp2.fr}

\author{David L. Lambert}
\affil{Department of Astronomy, University of Texas, Austin, TX 78712 USA;
dll@anchor.as.utexas.edu}

\author{Catherine A. Pilachowski} 
\affil{Astronomy Department, Swain West 319, 727 E. 3$^{\rm rd}$ St.,
Indiana University, Bloomington, IN 47405 USA;
catyp@astro.indiana.edu}

\author{Beatriz Barbuy}
\affil{Universidade de Sao Paulo, IAG, Av. do Matao 1226, Sao Paulo
05508-900, Brazil; barbuy@astro.iag.usp.br}

\author{Jorge Mel\'endez}
\affil{Universidad Nacional Mayor de San Marcos, SPACE \& IMCA, Lima,
Peru; jorge@astro.iag.usp.br}  

\author{Suchitra Balachandran}
\affil{Department of Astronomy, University of Maryland, College Park, MD
20742  USA; suchitra@astro.umd.edu}

\author{Michael S. Bessell}
\affil{Mount Stromlo \& Siding Spring Observatories, Institute of Advanced
Studies, The Australian National University, Weston Creek P.O., ACT 2611,
Australia; bessell@mso.anu.edu.au} 

\author{Douglas P. Geisler}
\affil{Departamento de Fisica, Universidad de Concepcion, Casilla 160 C,
Concepcion, Chile; doug@kukita.cfm.udec.cl}

\author{James E. Hesser}
\affil{Dominion Astrophysical Observatory, Herzberg Institute of
Astrophysics, National Research Council of Canada, 5071 West Sannich Rd.,
Victoria, BC V9E 2E7, Canada; jim.hesser@nrc.ca} 

\author{Cl\'audia Winge}
\affil{Gemini Observatory, Casilla 603, La Serena, Chile; cwinge@gemini.edu}

\begin{abstract}

High-resolution infrared spectra ($\lambda$/$\Delta$$\lambda$= 50,000)
have been obtained for twelve red-giant members of the Large Magellanic
Cloud (LMC) with the Gemini South 8.3m telescope plus Phoenix
spectrometer.  Two wavelength regions, at 15540\AA\ and 23400\AA, were
observed.  Quantitative chemical abundances of carbon (both $^{12}$C
and $^{13}$C), nitrogen, and oxygen were derived from molecular lines
of CO, CN, and OH, while sodium, scandium, titanium, and iron abundances
were obtained from neutral atomic lines.  The twelve LMC red giants
span a metallicity range from [Fe/H]= -1.1 to -0.3.  It is found that
values for both [Na/Fe] and [Ti/Fe] in the LMC giants fall below their
corresponding Galactic values (at these same [Fe/H] abundances) by about
$\sim$ 0.1 to 0.5 dex; this effect is similar to abundance patterns
found in the few dwarf spheroidal galaxies with published abundances.
The program red giants all show evidence of first dredge-up mixing of
material exposed to the CN-cycle, i.e. low $^{12}$C/$^{13}$C ratios,
and lower $^{12}$C- with higher $^{14}$N-abundances.  The carbon and
nitrogen trends are similar to what is observed in samples of Galactic
red giants, although the LMC red giants seem to show smaller 
$^{12}$C/$^{13}$C ratios for a given stellar mass.  This relatively
small difference in the carbon isotope ratios between LMC and Galactic
red giants could be due to increased extra mixing in stars of lower 
metallicity, as suggested previously in the literature.  Comparisons
of the oxygen to iron ratios in the LMC and the Galaxy indicate that
the trend of [O/Fe] versus [Fe/H] in the LMC falls about 0.2 dex below
the Galactic trend.  Such an offset can be modeled as due to an overall
lower rate of supernovae per unit mass in the LMC relative to the
Galaxy, as well as a slightly lower ratio of supernovae of type II
to supernovae of type Ia.

\end{abstract}


\clearpage

\section{Introduction}

The Large Magellanic
Cloud (LMC), one of the nearest galaxies
and a much smaller system than the Milky Way, is a prime target 
in which to probe chemical evolution
in stellar populations.  Unlike the Milky Way, where large
fractions of the volume are obscured by dust, the sightlines into
most of the LMC are relatively clear, rendering entire populations of
stars visible.
The LMC's distance demands that detailed
stellar abundance studies  
be based on spectra from 4-8 meter class
telescopes and be restricted to rather luminous stars.  

Pioneering abundance studies
have been conducted on LMC stellar samples already. 
One basic measure of overall chemical
evolution in a stellar system is an age-metallicity relation.  In
the LMC, Olszewski et al. (1991) and Dopita (1996) provide extensive
results, with the Olszewski et al. work based on clusters and Dopita's
results derived from planetary nebulae.  More recently, Geisler
et al. (1997) and Bica et al. (1998) have provided additional
results from clusters.  In summary, the LMC age-metallicity
relation appears to reflect a rapid enrichment phase more than 10 Gyr ago,
followed by apparent chemical quiescence until about 2 Gyr ago, when
another enrichment period (still ongoing) began.   

Additional insights into chemical evolution can be provided by 
studying
elements arising from different nucleosynthetic origins, e.g.
Type II supernovae (SN II), Type Ia supernovae (SN Ia), or asymptotic
giant branch (AGB) stars.  Studies of this type require high-resolution
spectra and are still a challenging endeavor to undertake in the LMC.
Published abundance distributions for the LMC include
Russell \& Dopita (1992), Barbuy, Pacheco, \& Castro (1994), Hill,
Andrievsky, \& Spite (1995), Luck et al. (1998), Korn et al. (2000),
Hill et al. (2000), Spite et al. (2001), and Korn et al. (2002).  The
analysed stars include main sequence and supergiant
B-stars,  supergiant K- to F-stars, and
Cepheid variables.  For the younger stars ($\le$ 10$^{8}$ yr), these
papers find a typical iron abundance of [Fe/H]$\sim$ -0.3 for
LMC field stars. 
The oxygen-to-iron abundance ratio in the LMC may be measurably
lower than this same ratio at the same [Fe/H] abundance in the Galaxy.
The trend
of [O/Fe] as a function of [Fe/H] can be a crucial relation in
establishing star formation histories in a stellar population.  Oxygen
is made preferentially in the most massive stars, while iron comes from
both massive, core-collapse supernovae (SN II) and the (presumably)
binary Type Ia supernovae.
The run of the O/Fe ratio with Fe/H in a stellar system is, therefore,
a measure of the history of SN II to SN Ia rates, and, hence, the
star formation history of  the LMC.
Carbon and nitrogen abundances may also be lower when compared 
to Galactic stars at
the same [Fe/H].  As C, N, and O (with Ne) are the most abundant
heavy elements,
their abundances carry much
weight in setting the overall metallicity, Z, in a galaxy.

Most of the previous abundance determinations of C, N and O in the LMC
have rested on atomic lines, some of which are quite weak, or hampered
possibly by non-LTE processes.  Because of the importance of the   O/Fe ratio
to the disentangling of the LMC's star fomration history,
this study set out to measure oxygen, as well
as carbon and nitrogen abundances, in a sample of LMC red giants by
observing molecular unblended lines of OH, CO, and CN from high-resolution
infrared (IR) spectra.   Using the 8.3m Gemini South reflector, along
with the Phoenix high-resolution IR spectrometer, abundances of the 
isotopes $^{12}$C, $^{13}$C, $^{14}$N, and $^{16}$O could be determined
along with abundances of Fe, Na, Sc, and Ti.

\section{Observations \& Data Reduction}

High-resolution infrared (IR) spectra were obtained of target red-giant
stars using the 8.3 m Gemini South reflecting telescope and the NOAO
Phoenix spectrometer (Hinkle et al. 1998; 2000; 2002).  
This instrument is a
cryogenically cooled echelle spectrograph that uses order separating
filters to isolate individual echelle orders.  The detector is a
1024x1024 InSb Aladdin II array.  The size of the detector in the
dispersion direction limits the wavelength coverage in a single
exposure to about 1550 km s$^{-1}$, or $\sim$120\AA\ at 2.3 $\mu$m
and $\sim$80 \AA\ at 1.6 $\mu$m.  One edge of the detector is blemished
so the wavelength coverage on all but the brightest sources is typically
trimmed a few percent to avoid this area.  The InSb material limits the
range of sensitivity to the 1-5$\mu$m wavelength region.  However, the
thermal brilliance of the sky requires that the observation of fainter
sources, e.g. LMC giants, be limited to the non-thermal 1-2.4$\mu$m
region.  The spectra discussed here were observed with the widest (0.35
arcsecond) slit resulting in a spectral resolution of
R=$\lambda$/$\Delta$$\lambda$= 50,000.

The program star positions, K-magnitudes, and (J-K) colors are listed
in Table 1.  The positions are taken from the
2-Micron All-Sky Survey (2MASS) database (http://irsa.ipac.caltech.edu). 
The first seven stars listed in Table 1 were selected based upon a 
color-magnitude diagram
constructed using the 2MASS database for two 15' fields in the LMC.
The LMC red giant branch is clearly visible in a K versus (J-K)
diagram and all program field giants were found to be LMC members based
on their radial velocities.   The identifying names given to the
field stars here were assigned by us as either field 1 or field 2,
followed by the star number from our list for that field, i.e. 2.4525
is from field 2 and is number 4525 from that list.  In addition to the 
field stars selected from the 2MASS survey, one star is a bright field 
supergiant (HV 2568), while four
stars in Table 1 were taken from Aaronson \& Mould's (1985) IR photometric
study of LMC clusters.  NGC1898 AM5 is found not to be a member
of the cluster (as will be seen from our radial velocity and abundance
analysis), but a foreground or background LMC giant.  
 
Each program star was observed along the slit at two, three, or four 
separate positions separated by 5" on the sky: the delivered image FWHM 
at the spectrograph varied from 0.25"-0.80" during the nights that spectra
were taken, so stellar images at different positions on the slit were
well separated on the detector. 
Equal integration times were used for a particular program star during
a particular set of observations.  With this observing strategy,
sky and dark backgrounds are removed by subtracting one integration
from another (the star being at different positions on the detector
array).  During each night, 10 flat-field and 10 dark images were
recorded for each given wavelength setting of the echelle grating.  A hot
star, with no intrinsic spectral lines in the regions observed, was also
observed each night in each observed wavelength region.  If telluric
lines were present in the spectral interval, the hot star was observed
at airmasses bracketing the program stars and suitable combinations of
the hot star spectra were used to divide out the telluric lines from 
program star spectra.  If no telluric lines fell in the observed region,
the hot star was observed only once as a monitor of fringing and general
test of data reduction quality.  All program stars were observed at two
echelle tilts through two different filters: one filter was K4308, with
wavelength coverage from 23345 to 23455\AA, and the other was H6420,
with wavelength coverage from 15505 to 15575\AA.  Table 2 provides
an observational log, showing the number and integration time for
each observation, as well as the approximate signal-to-noise (S/N)
of each final, reduced spectrum (with reduction as described below). 

The array image frames were reduced to one-dimensional, wavelength
calibrated spectra with the IRAF\footnote{The IRAF software is
distributed by the National Optical Astronomy Observatories under
contract with the National Science Foundation.} set of routines.  
For a given night
and wavelength setting, dark and flat-field frames were averaged 
and the dark frame was subtracted from the flat-field.  The average,
dark-subtracted flat-field was normalized and the pairs of differenced
program star frames were divided by the normalized flat-field.  The
aperture containing the stellar signal was defined and traced for
each star frame, followed by extraction of the spectrum.  For the
23400\AA\ spectra, wavelength calibrations were computed by using a
set of telluric wavelengths obtained from the hot star spectra.  In the 
15540\AA\ region, where there are no telluric lines, wavelengths were
set using photospheric lines from the red giants themselves, so no 
radial-velocity information can be extracted from the spectra observed
in this region.  The wavelength calibration fits yielded residuals of
typically 0.01--0.02\AA.    
 
Sample reduced spectra are shown in Figure 1 for both spectral regions 
in one of the field red giants (2.3256).  The top panel shows 
the 15540\AA\ region
and the bottom panel the 23400\AA\ region.  The 23400\AA\ region
contains a number of telluric absorption lines and this spectrum
has been ratioed with the hot star HIP27603 in order to remove
telluric absorption lines.  The effective airmass of the telluric
standard star is adjusted by differently weighted averages of the
different integrations on this star until the telluric absorption
cancels out as well as possible from the program star.  In practice
virtually no evidence of such absorption is visible in the program star
spectra.   In both regions, a number of the spectral lines or features 
used in the abundance analysis are identified.  This figure
illustrates both the quality of the spectra and the variety of
lines available for study.  

\section{Analysis}

The analysis uses a combination of
photometry and spectroscopy to determine fundamental stellar parameters
(effective temperature, surface gravity, microturbulent velocity, and
overall stellar metallicity) and detailed chemical compositions.  The
high-resolution spectra are analyzed using plane-parallel, LTE model
atmospheres that are generated using two different versions of the MARCS
code: an older version based largely upon the program as described in 
Gustaffson et al. (1975), and a more recent version, OSMARCS (Plez, Brett,
\& Nordlund 1992; Edvardsson et al. 1993; Asplund et al. 1997), which 
contains more extensive
molecular opacities.  Abundances derived from both types of models are
compared to give some indication of potential model atmosphere systematic
effects.  The model atmospheres are   
coupled to the most recent version of the LTE spectrum
synthesis code MOOG (Sneden 1973) in order to derive abundances.  

\subsection{Stellar Effective Temperatures}

The effective temperatures of the stars are chosen using broadband  
color (J-K)-T$_{\rm eff}$ calibrations.  Table 1 listed
the stars observed along with their K-magnitudes and (J-K) colors.
The 2MASS database is the source of the magnitudes
and colors for the field giants, while Aaronson \& Mould (1985) were
used for the cluster members.  The 2MASS magnitudes and colors have
been transformed to the system defined by Bessell and Brett (1988),
as we use T$_{\rm eff}$-calibrations from Bessell, Castelli, \& 
Plez (1998), who use the color system defined in Bessell \&
Brett (1988).  The 2MASS corrections are those defined by Carpenter (2001) 
in his Appendix A and are fairly small: a constant 0.04 magnitude
offset in K, and a small color term that is about 0.03 magnitude
in (J--K).  
Table 3 lists the absolute bolometric magnitudes based upon a distance
modulus of 18.5 and K-band bolometric corrections from
Bessell et al. (1998): no corrections for reddening in the IR were
included here.

Bessell et al. (1998) show 
that T$_{\rm eff}$-(J-K) relations  have a small metallicity
dependence and we use their calibration for [Fe/H]= -0.50, which is
what is expected for the typical LMC field (Cole, Smecker-Hane, \&
Gallagher 2000): Table 3 lists these effective temperatures. 
Other color effective temperature relations are available in the
literature. We compare one other source (McWilliam 1990) to the
calibrations used here.  The mean  difference between the effective
temperatures from the Bessell et al. (1998) calibration minus those
from the McWilliam (1990) calibration is
+90$\pm$51K,  suggesting that the temperature scales agree
at a level of about 100 K. 

\subsection{Surface Gravities, Microturbulent Velocities, \& Iron Abundances}

The stellar surface gravity 
can usually be derived
from either ionization equilibria (where, for example, Fe I
and Fe II lines are forced to yield the same abundance) or from estimates
of stellar luminosity and mass.  No ionized atomic lines are detectable in
the relatively small IR spectral windows observed here, so gravities
are defined by luminosity and mass estimates  from the
standard relation

 (g/g$_{\odot}$)= (M/M$_{\odot}$)$\times$(T$_{\rm eff}$/T$_{\rm eff}$($\odot$))$^{4}$$\times$(L$_{\odot}$/L).

 For  LMC red giants, bolometric magnitudes (Table 3) and, hence, 
absolute luminosities are known with some accuracy; the distance
modulus adopted is 18.5 $\pm$ 0.2. Mass estimates are obtained using
evolutionary tracks.
Figure 2 shows plots of M$_{\rm bol}$ versus T$_{\rm eff}$ for the LMC
red giants from this work and model tracks from Schaerer et al. (1993),
with the model tracks computed for the metallicity expected for most of
LMC program stars ([Fe/H]= -0.35).  Models are shown for 1.0, 2.0, 3.0,
4.0, 5.0, 7.0, and 10.0M$_{\odot}$ and these tracks stop 
at the tip of the first ascent of the
red giant branch (RGB).  Karakas \& Lattanzio (2002, private communication)
kindly provided tracks for 1.0, 1.9, 3.0, and 4.0M$_{\odot}$  
models that included evolution along the asymptotic giant branch (AGB) 
and into the thermally pulsing (TP-AGB) phase of evolution (with these
models also computed for an LMC metallicity).  The two separate sets
of models from Schaerer et al. (1993) and Karakas \& Lattanzio (2002)
agree well from evolution on the ZAMS and onto the RGB; on the AGB and
TP-AGB, the tracks become rather complicated and these tracks are not
plotted in Figure 2 for the sake of clarity. 

The top panel of Figure 2 provides a broad view of stellar evolution
and includes the positions of a few other samples of LMC stars from
previously published abundance studies, whose results will be compared 
to those derived here.  The Korn et al. (2002) paper is an analysis of 4
main-sequence B-star members of the young cluster NGC2004, while
Korn et al. (2000) studied 5 somewhat evolved B-star members in
two clusters (NGC2004 again and NGC1818).  Hill et al. (1995) derived
abundances in 9 field nonvariable (or very slightly variable) F-supergiants.
What Figure 2 makes clear is that compared to these earlier LMC studies,
the stars sampled here are of rather lower mass.
The earlier abundances mentioned above are derived from LMC
stars with masses of $\sim$10M$_{\odot}$, while the red giants analyzed
here have masses closer to $\sim$2-4M$_{\odot}$.  The mass and 
evolutionary regimes are quite different and thus our study can add
complementary abundance information to what has been derived
already for the LMC. 

The bottom panel of Figure 2 shows, in more detail, the  
exact location of the target stars from this study.  Except
for the massive supergiant, HV2586, the red giants fall in a narrow
region crossed by either first ascent giants, or early AGB stars with
masses in the range of $\sim$2-4M$_{\odot}$.  Individual masses were
estimated for each LMC red giant and these approximate masses are listed
in Table 3: these masses were obtained by comparisons to both the 
Schaerer et al. (1993) and Karakas \& Lattanzio (2002) tracks.
The star NGC1898 AM6 is worth a special note as it will
turn out to be significantly more metal-poor than the other stars in
this sample (with [Fe/H]= -1.13).  Model tracks with [Fe/H]= -0.35,
at a given mass, are too red for this giant and more appropriate
tracks indicate that it is a lower-mass giant with M$\sim$1.0M$_{\odot}$.
The surface gravities used for the abundance analysis were then calculated
from the mass estimates and are listed in Table 3, rounded to the nearest 
solar mass.  Since no program star, except the obviously
massive supergiant HV2586, appears likely to be above 
4M$_{\odot}$, and none apparently at masses lower than 2M$_{\odot}$
(except for the metal-poor cluster member NGC1898 AM6), it is 
likely that errors in the surface gravity do not
exceed $\sim$0.2 dex, which would not cause a significant error (i.e.,
$\ge$ 0.15 - 0.20 dex) in derived abundances.

Modeling the behavior of spectral lines with one-dimensional model
atmospheres requires the definition of the microturbulent velocity, $\xi$.
The value of $\xi$ is set by the condition that lines of differing
line-strengths from the same species result in the same abundances.
In the IR spectral windows observed here, the numerous $^{12}$C$^{16}$O,
hereafter $^{12}$CO lines in the 23400\AA\ window can be used, as the
various rotational lines from the (3-1) vibration series exhibit a
variety of line-strengths (see Figure 1).  The detailed selection of the 
$^{12}$CO lines is discussed in Section 3.4,
but Figure 3 shows an example of the determination of the microturbulent
velocity for the cluster-member red giant NGC2203 AM1.  The left panels
show values of slopes versus $\xi$ for the abundance versus
reduced equivalent-widths (top panel) and abundance versus excitation
potential slopes(bottom panel).  Both plots show a zero
slope at nearly the same value of the microturbulence, indicating that
the T$_{\rm eff}$ used for this star is a reasonably good value, as
the abundance versus excitation potential is sensitive to the 
temperature, thus, a unique value of $\xi$ provides a simultaneous
fit to both line-strength and excitation potential.  The right panels
illustrate the quality of the spectral-line data by showing the derived
abundances versus both reduced equivalent-width (top panel) and versus
excitation potential (bottom panel) for the best-fit value of
$\xi$= 3.3 km s$^{-1}$.  Both trends show effectively zero slopes
and small scatter ($\pm$0.05 dex in the $^{12}$C abundance).  The derived
microturbulent velocities are listed in Table 3.  

The right panels in Figure 3
also illustrate an experiment conducted on line formation in the model
atmospheres.  The $^{12}$CO lines are formed over a range of optical
depths in the atmosphere, with strong lines being formed preferentially
higher compared to weaker lines.  The filled circles in these panels
show results for a ``standard'' model atmosphere used in the abundance
analysis: standard here means the atmosphere extends to optical depths,
defined at $\lambda$=5000\AA\, as low as $\tau$$_{5000}$= 10$^{-4}$,
at which point the atmosphere is truncated.  The open squares show
abundances from the $^{12}$CO lines for a model that has been
truncated at $\tau$$_{5000}$= 10$^{-3}$: as some of the $^{12}$CO lines
are the strongest lines in the spectra, if substantial contributions to
their absorption were occurring near $\tau$$_{5000}$$\sim$ 10$^{-4}$,
truncating the models here would result in incorrect abundances.  Clearly,
truncating at either 10$^{-3}$ or 10$^{-4}$ in optical depth has no
significant effect ($\le$0.01-0.02 dex) on derived abundances, and the
optical depth structures of the models are adequate.

Earlier abundance studies of Galactic red giants that overlap somewhat
the effective temperatures and gravities of the stars studied here
were Smith \& Lambert (1985, 1986, 1990) and it is instructive to compare
values of the microturbulence derived from those papers and the
microturbulent velocities measured here; the Smith \& Lambert analyses
relied on near-IR Fe I lines as the primary indicators with which to 
set $\xi$, while using the $^{12}$CO lines as secondary checks.  
The top panel of Figure 4 is a plot of $\xi$
versus absolute bolometric magnitude for the LMC red giants and the
Galactic stars from the Smith \& Lambert papers.  It appears that
the lower luminosity giants have a typical $\xi$= 2.0 km s$^{-1}$,
but at luminosities greater than M$_{\rm bol}$$\sim$ -2.5, there is a
trend of increasing values of $\xi$ with increasing absolute magnitude,
as well as increasing amounts of scatter.  The LMC
red giants fall right within the scatter as defined by the Galactic
giants, with the LMC sample tending to be composed of stars of slightly
higher luminosities than the samples of Galactic stars studied by
Smith \& Lambert (1985, 1986, 1990), but exhibiting the same basic
behavior in terms of the microturbulence.  The bottom panel of Figure 4
is in a similar vein to the top panel: here $\xi$ is plotted versus
log g.  Microturbulent velocities for the Milky Way giants cluster
around 2.0 km s$^{-1}$ for the higher gravity stars (log g $\ge$ 1.0),
but then $\xi$ increases both in magnitude and scatter as gravity
decreases.  The overlap between Galactic and LMC giants is essentially
perfect. 

The metallicity is estimated from the 
15540\AA\
window where three Fe I lines are  measurable as largely
unblended features.  The next section covers the linelist selections
for the various species in detail, but the [Fe/H] values listed in
Table 3 are from these three lines, using gf-values set by fits to the
IR Arcturus spectra atlas (Hinkle, Wallace, \& Livingston 1995) 
using the
Fe abundance  determined from  optical Fe I and Fe II
lines (Smith et al. 2000).  The parameters for
$\alpha$ Boo (Arcturus) adopted here are those derived by Smith
et al. (2000): T$_{\rm eff}$= 4300K, log g= 1.7, $\xi$= 1.6 km s$^{-1}$,
and [Fe/H]= -0.72. 

The last column of Table 3 lists the heliocentric radial velocities
of the program stars as determined from the sample of nine
$^{12}$CO lines used to set the microturbulence.  From the radial
velocities it is found that all program stars are LMC members;
Prevot et al. (1985) find velocities of field stars in the LMC
to range from $\sim$+250 to +310 km s$^{-1}$.  The two clusters
in our sample were also observed by Olszewski et al. (1991) for
both radial velocities and metallcities (not the same stars as
observed here, however).  In the case of NGC2203, Olszewski et al.
(1991) find a mean radial velocity of +243 km s$^{-1}$, which is
close to the +253 km s$^{-1}$ found for the two member stars in
this sample.  They also derive a cluster metallicity of
[Fe/H]= -0.53: a value similar to what we find for AM1 (-0.67)
and AM2 (-0.61).  For NGC1898, Olszewski et al. (1991) observe a
mean cluster velocity of +210 km s$^{-1}$ and a metallicity of
[Fe/H]= -1.37.  This radial velocity and abundance are close
to what we find for AM6 (V= +219 km s $^{-1}$ and [Fe/H]= -1.1).
For the other star, AM5, we find V= +255 km s$^{-1}$ and 
[Fe/H]= -0.50: quite different from Olszewski et al. (1991)
or what we find for AM6.  This star is almost certainly a cluster
non-member, however, its velocity indicates that it is a member
of the LMC.  It is either a foreground or background LMC red 
giant in the field of NGC1898. 

\subsection{The Atomic Lines Used: Fe I, Na I, Sc I, and Ti I}

In the two spectral intervals sampled, a handful of atomic lines
are detectable which are largely unblended and single.
These lines are
listed in Table 4, along with the measured equivalent widths from the
twelve program LMC red giants.  The excitation energies are from
the Kurucz \& Bell (1995) lists, while the gf-values have been adjusted
to provide the same abundances from these lines in $\alpha$ Boo (Hinkle
et al. 1995) as found from the analysis of this standard star
in conjunction with analyses of $\omega$ Cen red giants (Smith et al. 2000).
Table 4 lists the line data and equivalent widths for all 12 of the
program stars. 

\subsection{The Molecular Lines Used for the CNO Abundances: CO, OH, and CN}

The IR windows were chosen to provide molecular lines from which to derive the
C, N, and O abundances. The 23400\AA\ region includes lines of the
CO first-overtone vibrationa-rotation bands. The 15540\AA\ region
includes lines of the OH first-overtone vibration-rotation bands
and  the CN Red system. 
Partial pressures of C and O are coupled through the formation of CO
and to a lesser extent of OH and H$_2$O. The partial pressure of N
is controlled almost exclusively by formation of N$_2$. 
Minor influences of other molecules was taken fully into account.
A simultaneous analysis of CO and OH lines provides the
C and O elemental abundances. Analysis of the CN lines and the C abundance
give the N abundance. 

The wavelengths, excitation potentials, and gf-values for the CO 
vibration-rotation lines are from Goorvitch (1994), with the
$\Delta$$\upsilon$=2 lines in the 23400\AA\ region being used.
The $^{12}$CO lines measured span a range in line strength and $\chi$,
and were used to set the microturbulence as well as provide a check on
the adopted T$_{\rm eff}$ from the (J-K) photometry.  The CO dissociation
energy was taken as D$_{0}$= 11.090eV.  In addition to the $^{12}$CO
lines, the $^{13}$CO (2-0) bandhead falls at the red edge of the
23400\AA\ setting in all but one program star, and synthesis of this
bandhead was used to derive a $^{13}$C abundance (tabulated as a
$^{12}$C/$^{13}$C ratio).   

The spectral region near 15540\AA\ was selected as it contains a number
of relatively isolated and unblended OH lines, as used in previous
oxygen analyses by Balachandran \& Carney (1996), Melendez, Barbuy,
\& Spite (2001), and Mel\'endez \& Barbuy (2002).  Asplund \&
Garcia P\'erez (2001) have pointed out that strong OH lines may be
sensitive to the effects of temperature differences and granulation
that are found in the 3D model atmospheres, but not in the 1D models;
however their calculations covered effective temperatures that are
considerably hotter (T$_{\rm eff}$= 5800-6200K) than those found here.
Nissen et al. (2002) point out that the granulation effects on the OH
lines from the 3D models are expected to decline with decreasing
effective temperature.  Mel\'endez \& Barbuy (2002) confirmed that
oxygen abundances derived from the stronger lines are higher than for
the weaker OH lines, again however, for stars that are distinctly hotter
than those sampled here.  The differences found by Mel\'endez \&
Barbuy (2002) are more pronounced for more metal-poor stars and disappear
for stars with [Fe/H]$\ge$ -1.  The combination of lower effective
temperatures and higher metallicities in the stars sampled here
why this effect is not found in our oxygen abundances.
The wavelengths, excitation potentials, and gf-values
are from Goldman et al. (1998), with the dissociation energy being
D$_{0}$= 4.392eV (Huber \& Herzberg 1979).

Nitrogen abundances are based on five CN lines in the 15540\AA\ spectral
region (these CN lines are weaker than the CO and OH lines, but are
well-defined and typically unblended).  The CN lines were chosen from
a linelist extracted from a masterlist compiled by B. Plez as described
in Hill et al. (2002).  The CN dissociation energy is D$_{0}$= 7.77 eV
(Costes et al. 1990). 

The molecular line data, as well as the measured equivalent widths are
listed in Table 4.

\subsection{Errors from Stellar Parameter Uncertainties and Final
Abundances}

As already discussed in Section 3.1, uncertainties exist in T$_{\rm eff}$
of at least $\sim$100K and gravities may be uncertain by about $\pm$0.2 dex.
Fits to the $^{12}$CO lines to determine $\xi$ constrain this parameter
to $\pm$0.2 to 0.3 km s$^{-1}$.  In order to investigate how these
parameter uncertainties manifest themselves into abundance changes,
the model for the program star 2.4525, which has an effective temperature
near the middle of the program star distribution, was perturbed by
+100K in T$_{\rm eff}$, +0.2 dex in log g, and +0.2 km s$^{-1}$,
respectively, and abundances computed for each perturbation.  These 
abundance changes are listed in Table 5 for each species used in
the abundance analysis.  These changes can be assumed to be quasi-linear
for small changes in the stellar parameters across the range of values
found for this sample of red giants and be used to assess the overall
uncertainties in the abundance analysis.  

Systematic effects are always
possible, for example, due to the choice of type of model atmosphere.
This possibility was investigated by comparing abundances derived
from two different sets of model atmospheres, one from the older
MARCS program and the other from the newer OSMARCS code.  In none
of the stellar parameters spanned by the program stars was it found
that derived abundance differences exceeded 0.10 dex.  In the deeper
layers (where $\tau$$_{5000}$$\ge$ 0.6) the two sets of models
agree in temperature and pressure.  In regions where spectral lines
form (typically $\tau$$_{5000}$$\sim$ 0.1), differences are still
essentially zero at T$_{\rm eff}$= 3800K, but the OSMARCS models become
slightly warmer (by about 40K), and have higher gas pressures (by about
20\%) for T$_{\rm eff}$= 3600K.  The lines affected most by the
differing models at this effective temperature were from OH 
(note the temperature sensitivity in
Table 5), however, the effects were small and of no significant
consequence to the abundances reported here.

Final abundances, based upon the OSMARCS models, are listed in 
Table 6 in the system of
A(x)= log $\epsilon$(x)= log[n(x)/n(H)] + 12.  
The $^{13}$C abundances are reported in the traditional form of 
$^{12}$C to $^{13}$C ratios.  For species with abundances 
derived from three or more lines, the standard deviation of the
mean is also shown: this provides some rough estimate of the
quality of the spectra used in the analysis.  All abundances were
based ultimately on spectrum synthesis, although the equivalent
widths listed in Table 4 were used for a first-cut abundance analysis.
Included in Table 6 are the adopted abundances in the Sun for the
elements studied, as well as for the well-studied, metal-poor red
giant $\alpha$ Boo.  The $\alpha$ Boo abundances for Fe, Na, Sc, and Ti
are those from Smith et al. (2000); the atomic-line gf-values were adjusted
so as to yield these abundances in $\alpha$ Boo from an analysis of
these lines in the IR Atlas spectrum of this star (Hinkle et al. 1995).
The abundances in $\alpha$ Boo for C, N, and O (including $^{13}$C) were
derived from the same spectral regions and lines of CO, OH, and CN
as used in the analysis of the LMC program stars. 

Examples of two spectral syntheses in LMC red giants are shown 
in Figure 5.  The top
panel shows one region containing three $^{12}$CO lines (of differing
strengths) and a Sc I line.  The spectrum displayed is that from
LMC 1.6 and three separate $^{12}$C and Sc abundances are illustrated
in the syntheses.  The comparisons between observed and model spectra
are excellent.  The bottom panel shows two of the OH lines in NGC1898 AM6
with, again, three different oxygen abundances in the synthetic spectra. 
Examples of observed and synthetic spectra of the blended
$^{13}$CO(2-0) bandhead are displayed in Figure 6.  The top panel
shows the LMC field giant 2.1158, which has an estimated mass of
$\sim$ 3M$_{\odot}$ (Figure 2 and Table 3).  Syntheses with three
different isotopic carbon ratios are shown, with $^{12}$C/$^{13}$C= 17
providing the best fit.  The cluster member AM6 of NGC1898 is 
shown in the bottom panel to illustrate $^{13}$CO in a lower-mass
($\sim$ 1 M$_{\odot}$), lower-metallicity ([Fe/H]= -1.1) giant.  In
this star, a much lower isotopic ratio is derived, with
$^{12}$C/$^{13}$C= 3.5. 

\subsection{The Abundances}

\section{Discussion}

Observational understanding of a stellar
system's chemical evolution 
must pursue a selection of elements whose
stellar surface abundances are considered to reflect
faithfully the composition of the gas from which the stars
formed. Here, Na, Sc, Ti, and Fe may plausibly be 
deemed such a selection.
Below, we discuss some implications of our abundances,
primarily, by contrasting abundance ratios relative to
Fe in the LMC, the Galaxy, and other systems.

On the other hand, surface
abundances of C and N in red giants
are well recognized to be affected by convective
mixing as red giants. This mixing rearranges the C and N
abundances and obscures the original abundances which
betray the chemical evolution of the stellar system.
We compare C and N in LMC and Galactic red giants.

Oxygen is likely unaffected by the convective mixing
until the giant evolves to the asymptotic giant
branch. Here, we assume O is a probe of chemical
evolution, and discuss the O/Fe ratio in the LMC
and the Galaxy.

\subsection{The LMC Iron Abundances}

The red-giant sample here consists of nine stars that are not associated
with any clusters and are thus categorized as field stars, and three stars 
that are members of clusters (1 in NGC1898 and 2 in NGC2203).  Recently,
Cole, Smecker-Hane, \& Gallagher (2000) have begun to map the field
metallicity distribution in the LMC, and present values of [Fe/H]
(derived from Ca II IR-triplet measurements) for 38 field red giants.
Their distribution for [Fe/H] is shown in the top panel of Figure 7.
This metallicity distribution peaks at about [Fe/H]= -0.5, with a
fairly sharp cutoff at -0.25, and a diminishing tail down to [Fe/H] of
about -1.6.  Our sample of field stars is small and the resulting [Fe/H]
distribution is shown in the bottom panel of Figure 7; clearly the
overlap in metallicity between the two samples is perfect.  The details of the
shape of the metallicity distribution function for the field stars studied
here cannot be well-defined with only nine stars, so a detailed comparison
with the Cole et al. (2000) results requires a larger sample, but the
overall agreement is encouraging.

The two clusters sampled in this study have both had their metallcities
measured by Olszewski et al. (1991), who also used the Ca II IR-triplet
method.  For NGC1898, they find [Fe/H]= -1.37, while we find
[Fe/H]= -1.13 for member AM5.  With estimated total uncertainties
of $\sim$0.16 dex in Fe I as derived from this study, plus the 
uncertainties from the Olszewski et al. (1991) work, a difference of
0.24 dex signifies quite good agreeement for this cluster.  
The other cluster
common to both studies is NGC2203, which has an average [Fe/H]= -0.64
from 2 members analyzed here (AM1 and AM2), while Olszewski et al. (1991)
find [Fe/H]= -0.53: again quite good agreement between two very different
types of analyses. 

\subsection{Sodium, Scandium,
and Titanium Abundances}

Commonly, results of stellar abundance analyses are expressed
as the quantity [X/Fe]. The Na, Sc, and Ti abundances in Table 6
give mean values [Na/Fe] = -0.30, [Sc/Fe] =  0.0, and
[Ti/Fe] =   -0.18 with a star-to-star scatter most probably
dominated by measurement errors. These results compare
fairly favorably with measurements from LMC  F-G supergiants and
Cepheids: for example, Hill et al. (1995) give ([Na/Fe], [Sc/Fe],
[Ti/Fe]) = (-0.2, -0.1, 0.0) for [Fe/H] $\sim$ -0.3 supergiants
referenced to the Galactic supergiant Canopus, and Luck et al. (1998)
give (-0.15, -0.14, 0.07) for similar Cepheids with respect to
Galactic Cepheids. Use of Galactic comparisons minimizes systematic
errors. At the same [Fe/H] as the LMC stars, Galactic stars
show slightly positive values of [Na/Fe] and [Ti/Fe] but a
[Sc/Fe] close to zero  At the same [Fe/H] as the LMC stars, Galactic stars
show slightly positive values of [Na/Fe] and [Ti/Fe] but a
[Sc/Fe] close to zero 
The mild Na and Ti underabundances of the LMC stars relative to
Galactic stars of the same [Fe/H] is likely attributable to the
same factors responsible for the lower [O/Fe] (see below).  In
addition, the underabundances of Na and Ti, relative to Fe, are
also found in the few dwarf spheroidal galaxies with published
abundances, e.g. Shetrone, Cote, \& Sargent (2001).

\subsection{Carbon and Nitrogen Abundances}

Carbon and nitrogen abundances are altered during mixing in red giants,
as material exposed to the CN-cycle is brought to the star's
surface.  The $^{12}$C abundance decreases, the $^{14}$N abundance
increases, and the $^{12}$C/$^{13}$C ratio decreases, but the
sum of the $^{12}$C, $^{13}$C, and $^{14}$N is conserved.  This
mixing in first ascent and early-AGB red giants is not expected
to alter the $^{16}$O abundance.

Figure 8 investigates the $^{12}$C and $^{14}$N abundances in the
LMC red giants in comparison to other samples of stars, both in
the LMC and in the Galaxy.  The abundances of $^{14}$N are plotted
versus the abundances of $^{12}$C.  The solar symbol
is located at A$^{12}$C=8.41 (Allende Prieto, Lambert, \& Asplund 2002--from
their 1D MARCS model) and A$^{14}$N=7.93 (Holweger 2001) and the dashed line
represents equal changes in both carbon-12 and nitrogen-14.  The solid
curves are various ``mixing curves'' in which initial abundances of
carbon and nitrogen are changed in such a way that their sum remains
constant: as $^{12}$C decreases, $^{14}$N increases to mimic red-giant
mixing of material consisting of increasing exposure to the CN-cycle.
The open circles show the
$^{12}$C and $^{14}$N abundances in Galactic disk M-giants from
Smith \& Lambert (1985; 1986; 1990) and these giants are well-represented
by a CN-cycle mixing curve defined by approximate solar carbon and
nitrogen initial abundances.  The four-pointed crosses are s-process
enriched MS and S stars, also from the Smith \& Lambert papers and
are plotted as different symbols because the photospheres of these giants
have presumably been contaminated by material from the third dredge-up,
associated with $^{4}$He-burning thermal pulses along the AGB.  Such
stars quite possibly have had primary $^{12}$C added to their surfaces
and might be expected to evolve away from a simple CN-cycle mixing curve.
What should be noted concerning the MS/S stars, however, are the
few examples which show very large $^{14}$N abundances.  These objects
are AGB stars that have undergone hot-bottom burning (e.g. D'Antona \&
Mazzitelli 1996),
where some amount of primary $^{12}$C produced during thermal pulses
on the AGB is converted to $^{14}$N, via the CN-cycle, at the base
of a deep convective envelope. 

The LMC red giants analyzed here are plotted in Figure 8 as filled circles
and fall in a much different region than the Galactic red giants.
The two mixing curves that define approximately the region in the
carbon-nitrogen plane where the LMC giants fall begin at lower initial
carbon and nitrogen abundances.  As initial values, we used the recent
paper by Korn et al. (2002) that presents abundances for 4 true
main-sequence B-star members of the young LMC cluster NGC2004.  They
conducted a non-LTE abundance analysis and their results for C and N
are shown as the open squares in Figure 8.  Using initial C and N
abundances that bracket the Korn et al. (2002) abundances, the 
resulting CN-mixing curves provide excellent approximations to  
the behavior of the field LMC red giants.  One of the results for LMC
initial abundances, as indicated by Korn et al. (2002), was a lower value
for both [C/Fe] and [N/Fe] in the LMC: such mixing lines, beginning
at values of $\sim$-0.3 dex in [C/Fe] and [N/Fe], go right through
the regions defined by the red giant abundances.  An additional
conclusion from the comparison of the carbon and nitrogen abundances
in both the main-sequence B-stars and the red giants, is that simple
CN mixing does an excellent job of describing the dredge-up in this
sample of red giants; no evidence of deeper mixing, or of hot bottom
burning, is found: these processes might lower the star's $^{16}$O
abundance through conversion of some $^{16}$O to $^{14}$N through
the ON-cycles.  

The constraint on possible oxygen depletions can be
strengthened further by investigating the C/N abundance ratios.  Most
of the LMC giants in Figure 8 fall at values of [$^{12}$C/Fe] that
are about 0.6 to 0.8 dex below their probable initial values (as
suggested by the Korn et al. 2002 results).  At these levels of carbon-12
depletions, for the case of only CN-cycle mixing, the expected C/N
ratios would be about 6 or less; all of the measured C/N ratios 
calculated from the abundances in 
Table 6 fit this constraint.  The C/N ratio will increase over this
value if additional nitrogen is added from the conversion
of $^{16}$O to $^{14}$N via the ON cycles.  Because oxygen is the most
abundant member of the CNO trio, even a relatively modest depletion
will measurably increase the nitrogen abundance.  Given a plausible
initial oxygen abundance as determined by Korn et al. (2002), a 
decrease in $^{16}$O of 0.2 dex would increase the C/N ratio to
values of 8-10 in the regime of carbon depletions observed in the
present sample of LMC red giants; such a large ratio is not found
in any of the program stars.   
The LMC red giants studied here appear to have effectively unaltered
$^{16}$O.  

The $^{12}$C/$^{13}$C ratio is another abundance indicator that is
sensitive to stellar mixing on the red giant branch.  The initial
ratio is expected to be quite large ($\sim$ 40-90), while first
dredge-up lowers this ratio to $\sim$ 18-26 for stars in the mass
range from M$\sim$ 1-8M$_{\odot}$ according to standard stellar
models (e.g., Charbonnel 1994; but also Iben 1964).  It has been
known for some time that low-mass field giants exhibited lower
$^{12}$C/$^{13}$C ratios than predicted by standard models, with Keller,
Sneden, \& Pilachowski (2002) providing recent results of a discussion
of this effect.  Masses for giant-star members of clusters can be
determined more accurately than masses for field giants, and the
effect of lower $^{12}$C/$^{13}$C ratios in lower-mass giants
has been quantified by Gilroy (1989) using open clusters.  Her
results of $^{12}$C/$^{13}$C versus giant mass are plotted in
Figure 9 as the open squares.  Above a mass of about
2.0-2.5M$_{\odot}$, the values of $^{12}$C/$^{13}$C are 22-30 and
agree reasonably well with standard stellar models of first
dredge-up.  In the lower-mass giants, however, Gilroy (1989)
finds steadily decreasing $^{12}$C/$^{13}$C ratios with decreasing
mass, in contradiction to standard models.  The $^{12}$C/$^{13}$C
ratios derived here for the LMC red giants are also plotted in
Figure 9 as the filled circles.  The masses for the LMC giants
are estimated to only $\pm$0.5M$_{\odot}$ and are based on the model
tracks from Figure 2: these masses should be viewed as crude estimates.
Nevertheless, there is a general trend of lower $^{12}$C/$^{13}$C
ratios with lower masses that tracks the Gilroy (1989) trend
rather well, although offset to lower isotopic carbon-12 to -13
ratios for a given mass.  Charbonnel (1994) and Charbonnel, Brown,
\& Wallerstein (1998) investigate the lower $^{12}$C/$^{13}$C
values in the lower-mass red giants and suggest that extra-mixing,
possibly driven by meridional circulation and rotationally induced
turbulence can explain the lower $^{12}$C/$^{13}$C ratios.  In
their model, extra-mixing is inhibited by mean molecular weight
gradients, which increase with increasing metallicity, so they
predict more extensive mixing (and lower $^{12}$C/$^{13}$C ratios
for a given-mass giant) at lower metallicities.  The offset between
Gilroy's (1989) trend and the LMC trend may be due to this 
metallicity effect.   

The $^{12}$C/$^{13}$C ratio of 45 in NGC2203 AM2 deserves some
mention.  Although the errorbars on this measurement are large,
it still falls well above the trend of $^{12}$C/$^{13}$C with
mass found for the other LMC red giants.  One possibility is
that this star has begun dredge-up as an AGB star and some
$^{12}$C (produced from $^{4}$He-burning) has been added to the
photosphere.  If the $^{12}$C abundance were increased by 0.15
dex due to such dredge-up, the observed current $^{12}$C/$^{13}$C
ratio of 45 would have been 30 before the addition of extra
$^{12}$C: within the scatter of the other red giants.  Until more
LMC red giants are observed, such an explanation of the
anomalous position of NGC2203 AM2 in Figure 9 is speculation.

\subsection{The Oxygen Abundances}

On the assumption that the oxygen abundance is left unaltered
by the convective mixing, an assumption supported by the
$^{12}$C, $^{13}$C, and $^{14}$N abundances of the LMC red
giants, the O/Fe ratio may be used as a tracer of chemical
evolution.  The [O/Fe] values run from -0.3 to +0.3 with a
hint that they increase with decreasing [Fe/H]. 
Oxygen to iron abundances from a large number of Galactic and
LMC studies, presented as [O/Fe] versus [Fe/H], are brought
together in Figure 10: the LMC red-giant results from this 
study are plotted as filled circles.
The Galactic samples of field stars shown in this figure
are taken from Edvardsson et al. (1993), shown as
open circles, Nissen \& Edvardsson (1992) shown as open triangles, 
Cunha, Smith, \& Lambert (1995) shown as open hexagons, and Smith,
Cunha, \& King (2001) shown as open squares (the Galactic results
are plotted as smaller symbols simply to keep these numerous points
from falling on top of each other and obliterating their visibility).
The open pentagons at low metallicity are the recent results for Galactic
stars from Melendez \& Barbuy (2002) using the IR OH v-r lines: this
paper also includes the earlier results from Melendez, Barbuy, \&
Spite (2001).
Other low-metallicity results for [O/Fe] are from Barbuy (1988) and
Nissen et al. (2002), shown as asterisks and three-pointed stars,
respectively.
Additional [O/Fe] values for the LMC are shown as comparisons to the
red giants observed here: the filled triangles are from Korn et al. (2002)
for main-sequence B-stars, the filled squares are from Hill et al. (1995)
for F-supergiants, and the filled pentagons are from Hill et al. (2000)
and Spite et al. (2001) for old clusters and globular clusters.  
Figure 10
shows that, over the metallicity range of [Fe/H]$\sim$ -1.0 to -0.2, the
LMC stars fall about 0.2-0.3 dex below stars in the Milky Way.  The
fact that main-sequence B-stars, F supergiants, and cool, red giants
all yield the same results from independent analyses using very different
sets of lines (O II and Fe III for B stars, weak O I 6158\AA\ lines
and Fe II for F supergiants, and Fe I plus OH for the red giants) suggests
strongly that this is now a well established trend.  

The common interpretation of the different [O/Fe] - [Fe/H] relations
is that star formation in the LMC occurred in an early burst.
Interruption of star formation reduced the input of oxygen into the
interstellar gas.  In this interval, SN Ia delivered iron and so
brought the O/Fe down.  In the Galaxy, star formation was less
affected by bursts, and O/Fe remained at its ``high'' value to
a higher [Fe/H] than in the LMC. 
The solid curves shown in Figure 10 are a simple interpretation of the
differing values of [O/Fe] in the Milky Way versus the LMC.  These
curves were generated by taking yields from supernovae of types
II and Ia and adding the processed elements into a unit mass of gas
over time.  The rate at which the abundances grow with time depends
on the supernovae rate per unit mass of gas.  This rate is normalized
by assuming a rate of one SN II per 100 years and one SN Ia per 300
years in the Milky Way, with the mass taken to be 
10$^{11}$M$_{\odot}$.  Oxygen mass yields for SN II were taken from
Arnett (1999) and convolved with a Salpeter mass function to yield an
average mass of 1.5M$_{\odot}$ ejected per SN II.  Iron yields for SN II
were similar to that from Timmes, Woosley, \& Weaver (1995), and set
at 0.15M$_{\odot}$ per SN II.  With the rates and yields set for the
Milky Way SN II, it remained to set both the Fe yield from SN Ia and
the time delay for their initial contribution to chemical evolution.
In order to fit the Milky Way trend as shown in Figure 10, it was
necessary to set the SN Ia Fe yield to 0.7M$_{\odot}$ per SN Ia and
to use a time delay of 1.5 Gyr: both of these values are reasonable
in such a simple approximation.  The resulting curve does a fairly
good job of following the values of [O/Fe] versus [Fe/H] in the
Milky Way, with the final point on the calculated curve shown being
reached after a time of 12 Gyr.  The other curve shown was generated as
an approximation to the LMC trend of [O/Fe] versus [Fe/H].  Although not
a unique solution, this curve was generated by simply lowering the
SN II rate per unit mass in the LMC by a factor of 3 and the SN Ia
rate by a factor of 2 (the SN II and Ia O and Fe yields and SN Ia
time delay were the same).  This very simple numerical model indicates
the [O/Fe] abundance ratios in the LMC can be understood as arising
from both a slower rate of SN II per unit mass (caused presumably by
a lower average star formation rate per unit mass of gas) and a slightly
enhanced ratio of SN Ia rates to SN II rates in the LMC.  These results
are in rough agreement with those suggested by the analytical chemical
evolution models of Pagel \& Tautvaisiene (1998).   

\section{Conclusions}

Abundances of seven elements (C, N, O, Na, Sc, Ti, and Fe) have been
measured in twelve red-giant members of the LMC from high-resolution
IR spectra obtained with Gemini South plus Phoenix.  Using IR spectra it
is possible to extract quantitative chemical abundances for a number of
elements in somewhat lower-mass stars (M$\sim$ 2-4M$_{\odot}$ red giants)
in the LMC than previous high-resolution optical spectroscopic studies
of main-sequence B-stars or F to K supergiants (with M$\sim$ 
8-10M$_{\odot}$).  The IR abundance analyses add complementary stellar
mass targets to the earlier works.  In addition, the molecular lines
of such species as CO, OH, or CN in the IR spectra allow for the
determination of the abundances of specific CNO isotopic species, such
as $^{12}$C, $^{13}$C, $^{14}$N, and $^{16}$O here, or $^{17}$O,
$^{18}$O, and $^{15}$N in future studies.

The iron abundances sampled here range from [Fe/H]= -1.1 to -0.3.  Both
[Na/Fe] and [Ti/Fe] are found to be consistently lower than their
Galactic values by $\sim$ -0.1 to -0.5 over the metallicity range
sampled in the LMC.  These characteristic underabundances of Na and
Ti seem to also occur in a number of dwarf spheroidal galaxies
(Shetrone et al. 2002).  

The LMC red giants in this sample all show evidence
of the mixing of CN-cycle material to their surfaces via the first
dredge-up, with $^{14}$N enhanced by +0.4 to +0.8 dex over its 
estimated initial
values, and $^{12}$C decreased by -0.3 to -0.5 dex.  No evidence is found,
in these predominantly first ascent giants or early AGB stars, of the
extreme nitrogen enrichments ($\sim$ +1.0 dex or more) that might result
from second dredge-up or hot bottom burning.  The $^{12}$C/$^{13}$C
ratios in the LMC red giants are found to decrease with decreasing giant
star mass in a manner similar to that found for Galactic red giants
(Gilroy 1989); however, the LMC trend appears to be shifted to lower 
$^{12}$C/$^{13}$C ratios for a given red-giant mass.  This shift may
be due to the increased mixing associated with lower-metallicity giants
as suggested by Charbonnel et al. (1998). 

A comparison of [O/Fe] versus [Fe/H] between the LMC and the Milky Way
finds that the LMC trend falls below by about 0.2 dex over the
range of [Fe/H]= -1.1 to -0.3.  Good agreement in [O/Fe] as derived
from a sample of F-supergiants (Hill et al. 1995), main-sequence B-stars
(Korn et al. 2002), and the red giants analyzed here suggests that
the difference between the LMC and the Milky Way is real.  Lower values
of [O/Fe] in the LMC can be explained by both a lower supernovae rate
(caused by a lower star formation rate) and a lower ratio of supernovae
type II to supernovae type Ia.

We thank the staff of Gemini South for their excellent assistance with
these observations.  Nick Suntzeff is to be thanked for insightful
comments concerning the distance to the LMC. 
This work is supported in part by the National Science Foundation through
AST99-87374 (V.V.S.).

\clearpage

\figcaption[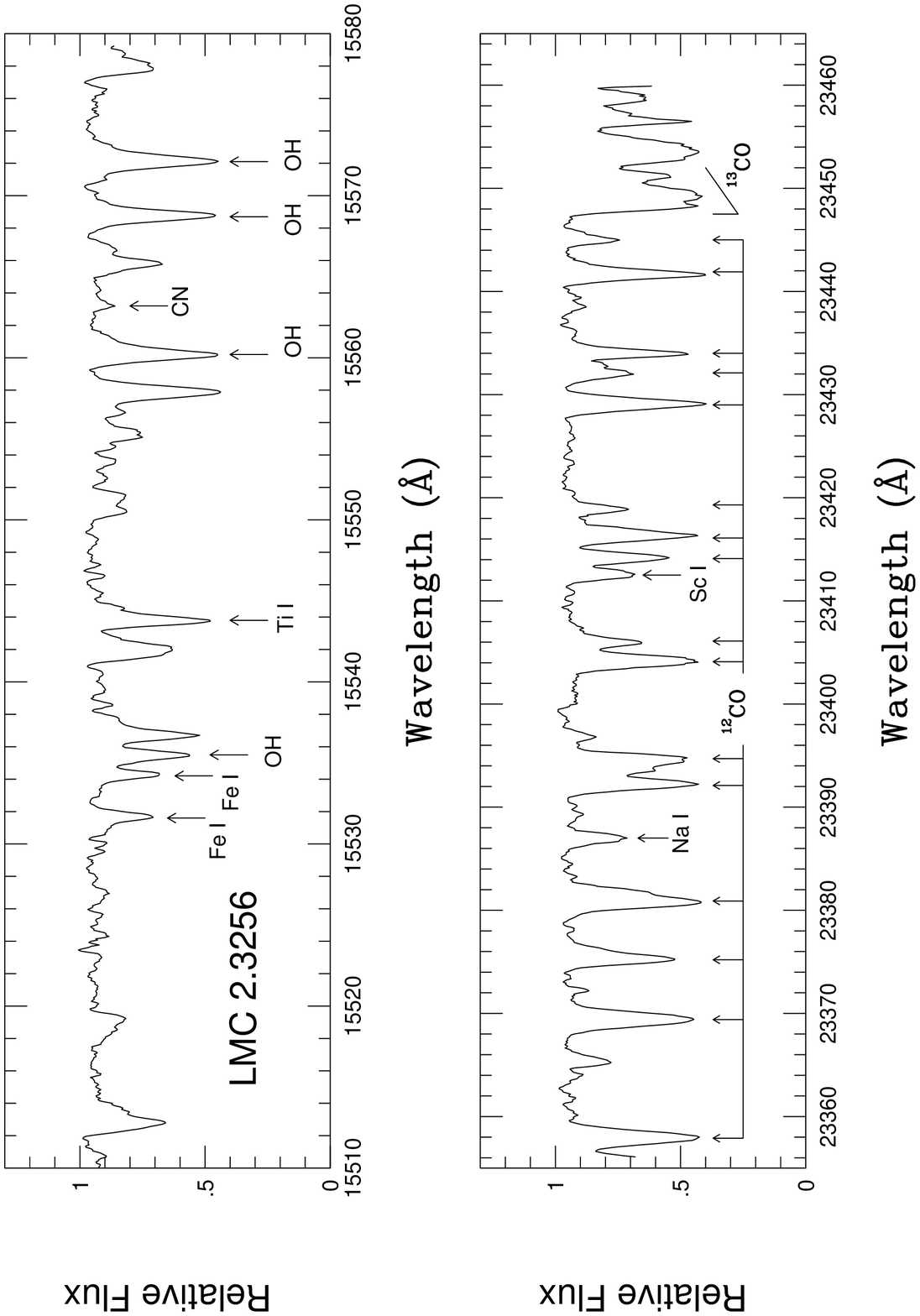]{Sample spectra of the LMC red giant 2.3256 (K= 12.0)
with each panel showing a different echelle grating tilt.  
Some of the stronger lines
are identified.  Each spectrum consists of a combination of three 
1/2-hour integrations and the signal-to-noise in each spectrum is a bit
more than 100 (see Table 2).  These spectra are representative of the
overall quality of the data analyzed here. 
\label{fig1}}

\figcaption[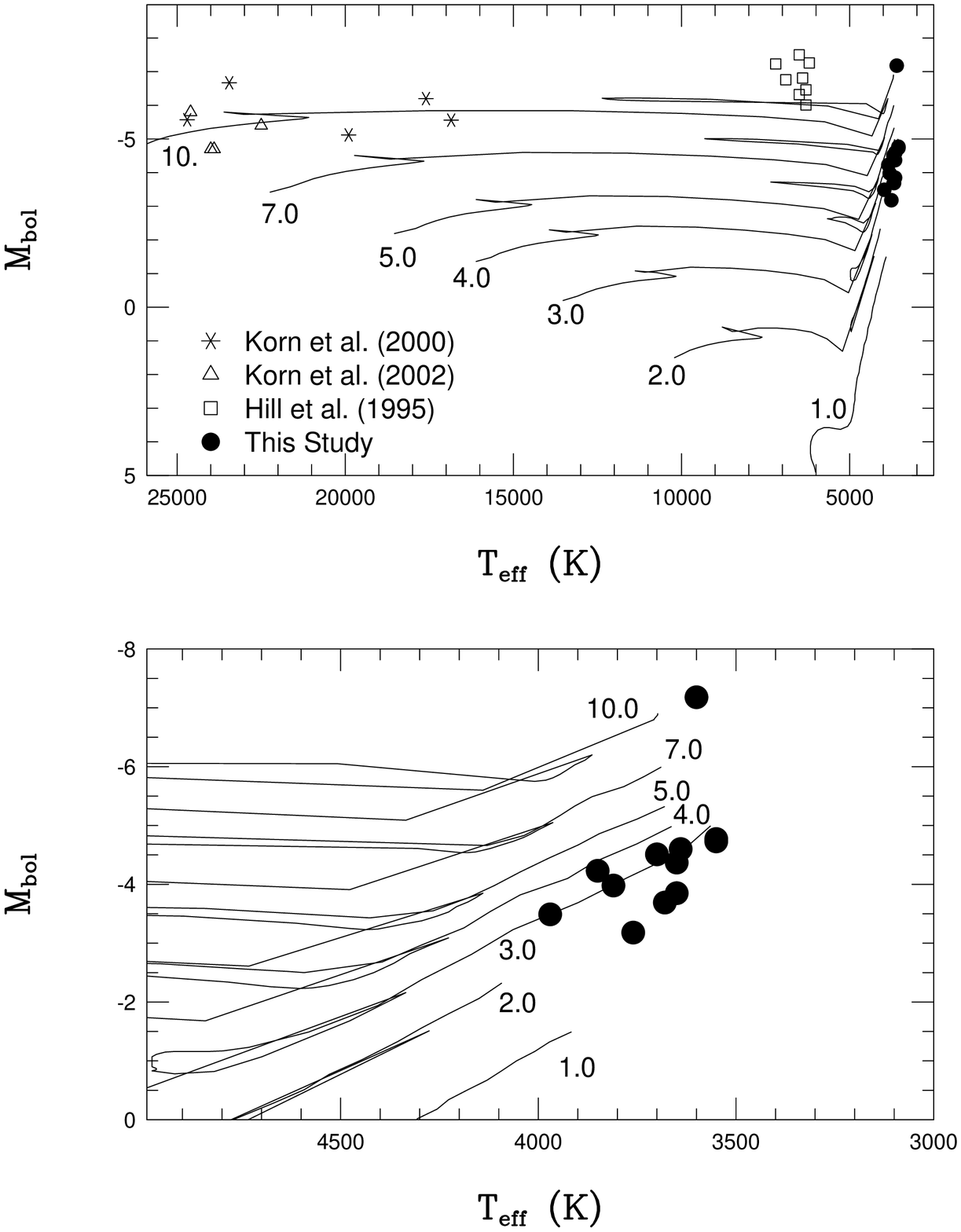]{Modified Hertzsprung-Russell diagrams, with absolute
bolometric magnitude versus effective temperature, for the red giants
analyzed here, a few other LMC stellar samples with published abundances,
and stellar model tracks from Schaerer et al. (1993).  Note that
previous LMC stellar abundance analyses tended to sample the more massive
stars, while the red giants presented here have lower masses.
The bottom panel shows an expanded view of the red giant region, with the
model tracks terminated at the tip of the red giant branch.  Models of
AGB stars were computed by Karakas \& Lattanzio (2002, private 
communication), however their tracks are complex and are not plotted
here for clarity.  The AGB tracks confirm that the red giants in this
sample are either first ascent stars or early-AGB stars.  Mass estimates
for these LMC giants are based on the Schaerer et al. tracks and span a
narrow range of $\sim$2-4M$_{\odot}$, except for the massive supergiant
HV2586 and the lower-mass metal-poor giant NGC1898 AM6. 
\label{fig2}}

\figcaption[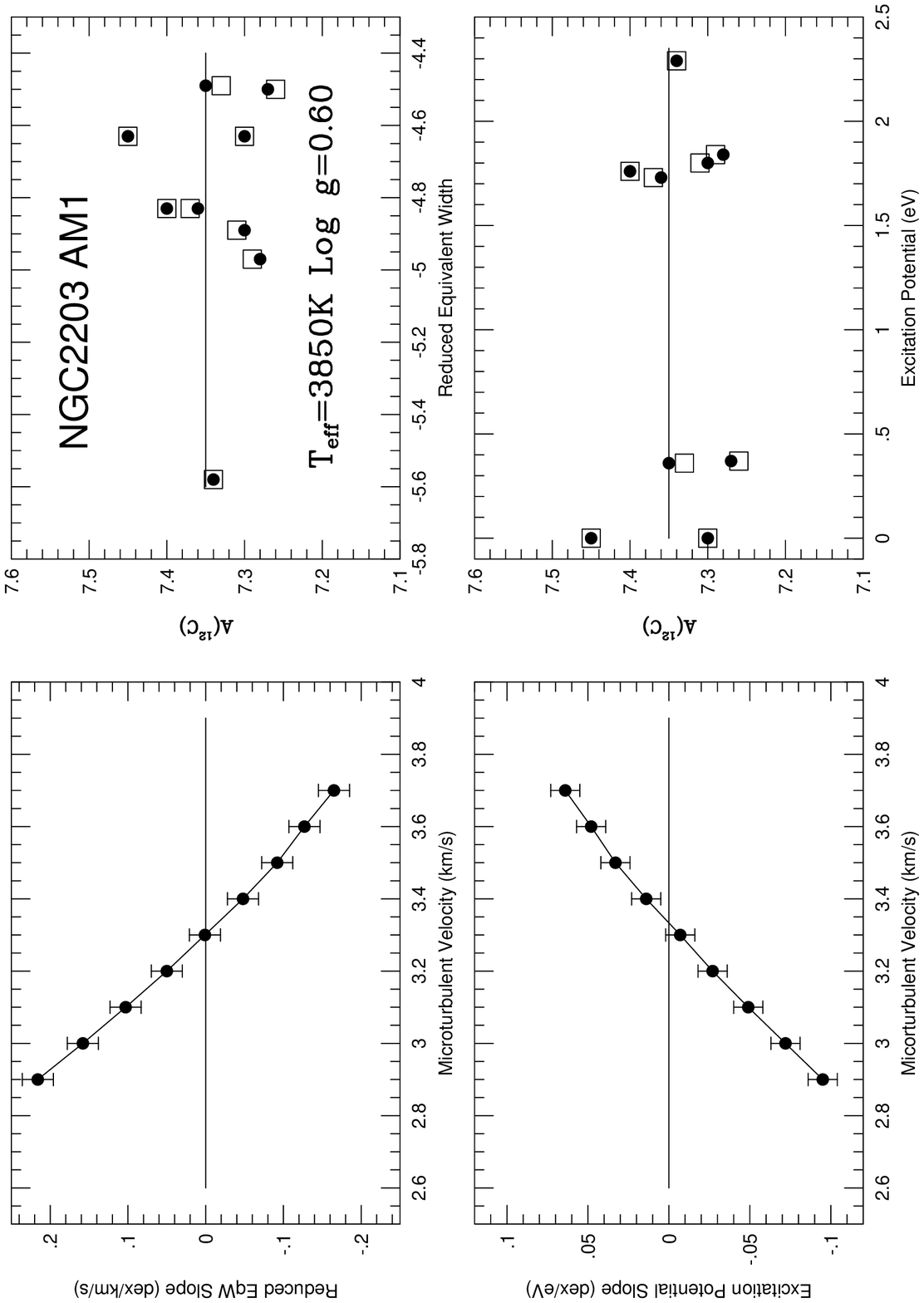]{An illustration of the determination of the
microturbulent velocity, $\xi$, for one red giant (AM1 in NGC2203).
The left panels show the linear slopes resulting from fits to 
the carbon abundances (derived from CO lines) versus reduced equivalent 
width (top) and versus excitaiton potential (bottom) for various
microturbulent velocities.  No trend, i.e. zero slope, for both reduced
equivalent width and excitation potential signifies a best fit.  The
effective temperature was set by IR colors, and note that a single
value of $\xi$ also results in a zero slope with excitation potential
(which is most sensitive to T$_{\rm eff}$).  The simultaneous
solutions for both reduced equivalent width and excitation potential
indicates a satisfactory fit for both T$_{\rm eff}$ and $\xi$.  The
right panels show the quality of the abundances for the given 
T$_{\rm eff}$-$\xi$ solution (a scatter of about $\pm$0.06 dex) for
this star.  The filled circles result from a model atmosphere extending
to an optical depth at 5000\AA\ of $\tau$= 10$^{-4}$ (what we usually
use as the upper atmospheric cut-off), while the open squares are
the abundances resulting from truncating the atmosphere at $\tau$=
10$^{-3}$.  As some of the $^{12}$CO lines are the strongest ones used,
their insensitivity to model atmosphere details in the regime $\tau$=
10$^{-3}$ to 10$^{-4}$ indicates that the atmospheric structure is
adequate for this abundance analysis.
\label{fig3}}

\figcaption[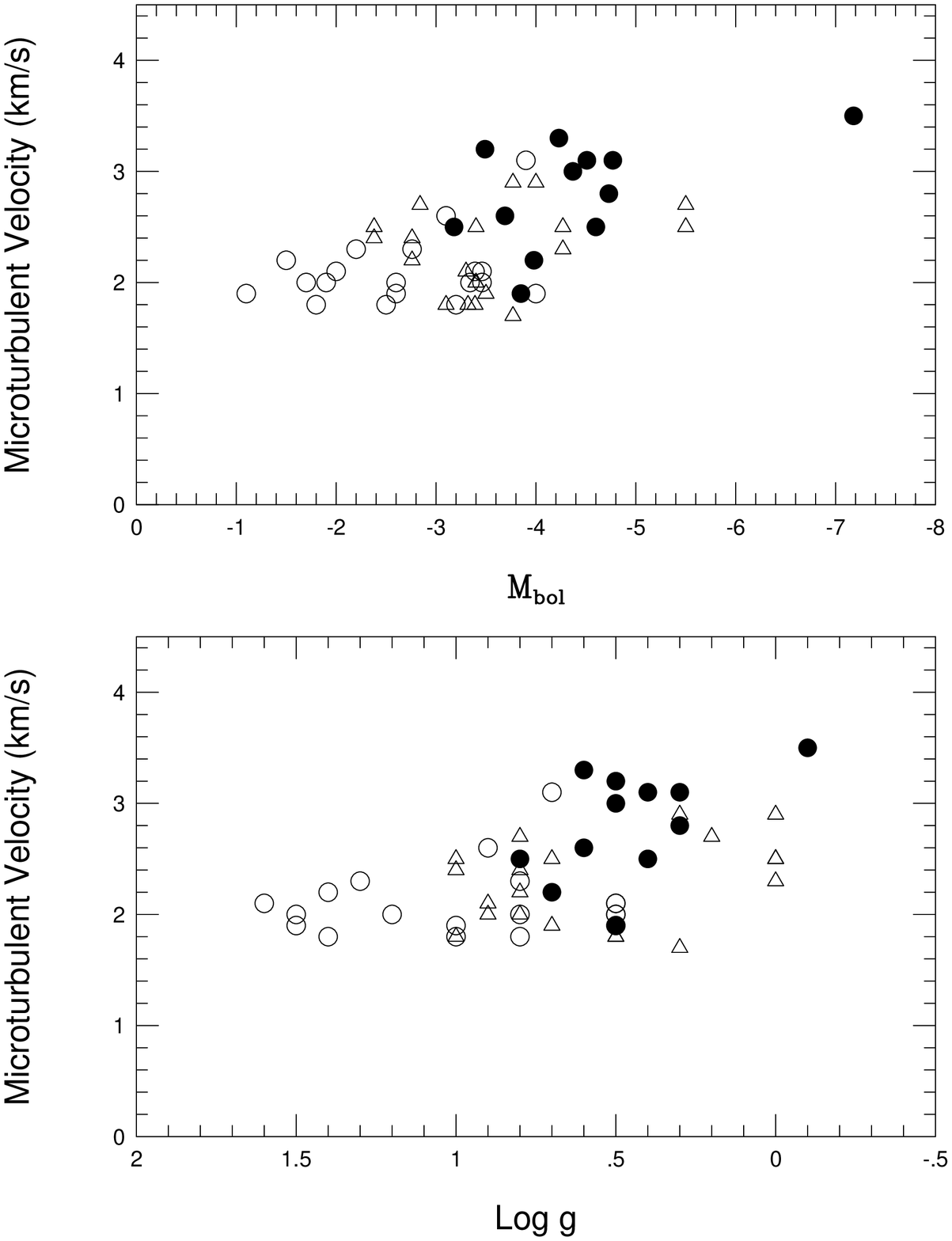]{Microturbulent velocities, as determined for Galactic
M-giants (open circles), or MS- and S-stars (open triangles) from
Smith \& Lambert (1985; 1986; 1990), and the LMC red giants (filled
circles) versus absolute bolometric magnitude (top panel) and logarithm 
of the surface gravity (bottom panel).  The overlap in $\xi$ with both
M$_{\rm bol}$ and log g between the Galactic giants and the LMC giants
is excellent.  At lower luminosity (with M$_{\rm bol}$$\ge$ -3)
and larger log g ($\ge$ 1), there is an almost
constant microturbulence of $\xi$$\sim$ 2.0-2.5 km s$^{-1}$.  As 
luminosity increases and surface gravity decreases, $\xi$ increases
slowly towards 3.0-3.5 km s$^{-1}$, with the Galactic and LMC giants
following a similar trend. 
\label{fig4}} 

\figcaption[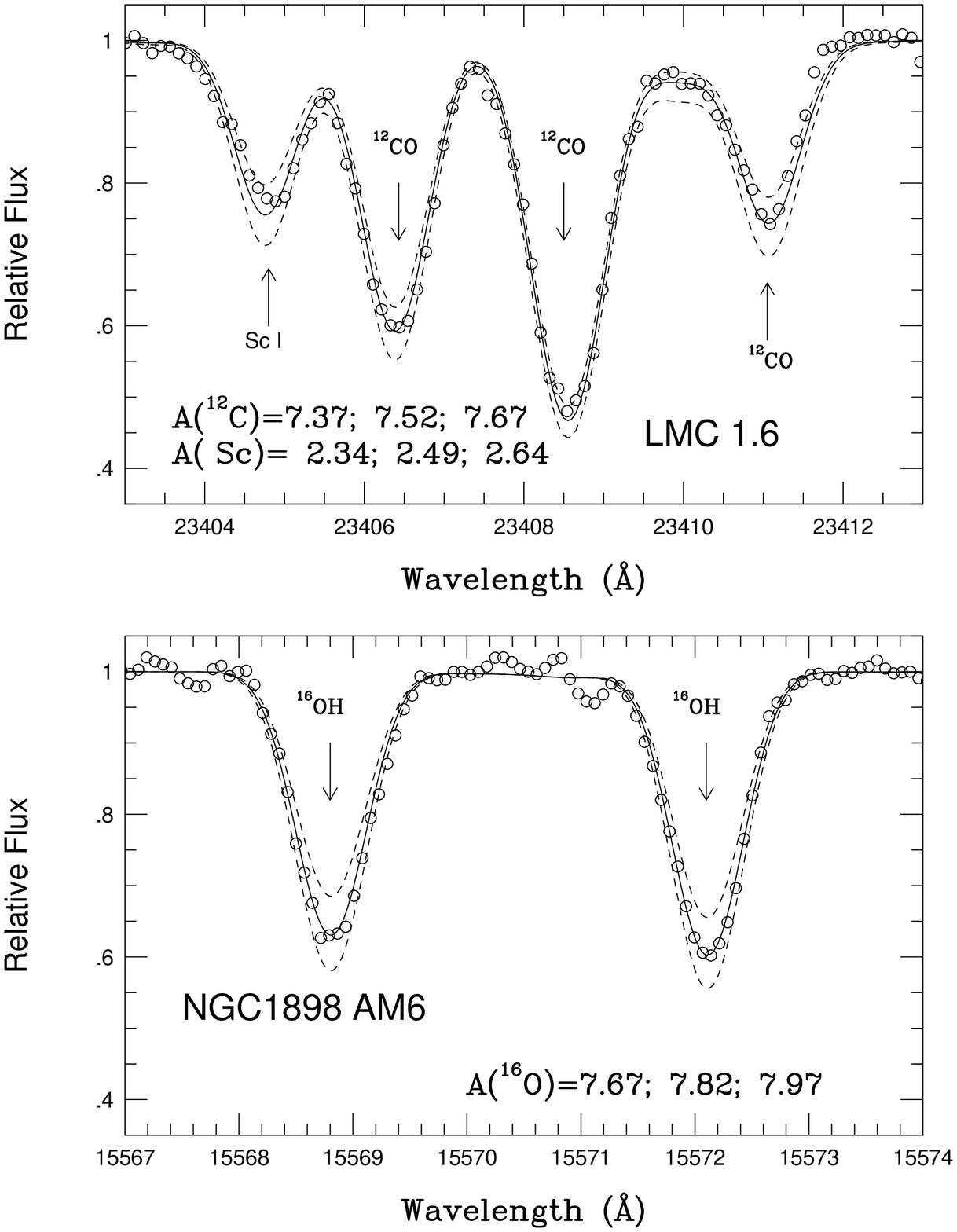]{Sample synthetic and real spectra in two spectral
regions in two different stars.  The top panel illustrates a small 
spectral region in the K-band showing three $^{12}$CO lines, of differing
strengths, and an atomic Sc I line.  The open circles are the observed
data, while the continuous curves are synthetic spectra with differing
carbon and scandium abundances (as shown on the figure).  Note that the
three $^{12}$CO lines, with very different equivalent widths and 
excitation potentials, yield nearly the same $^{12}$C abundance.  The
bottom panel shows two OH lines in the H-band, again with the open
circles being the real spectrum and the continuous curves showing 
synthetic spectra with differing $^{16}$O abundances.
\label{fig5}}

\figcaption[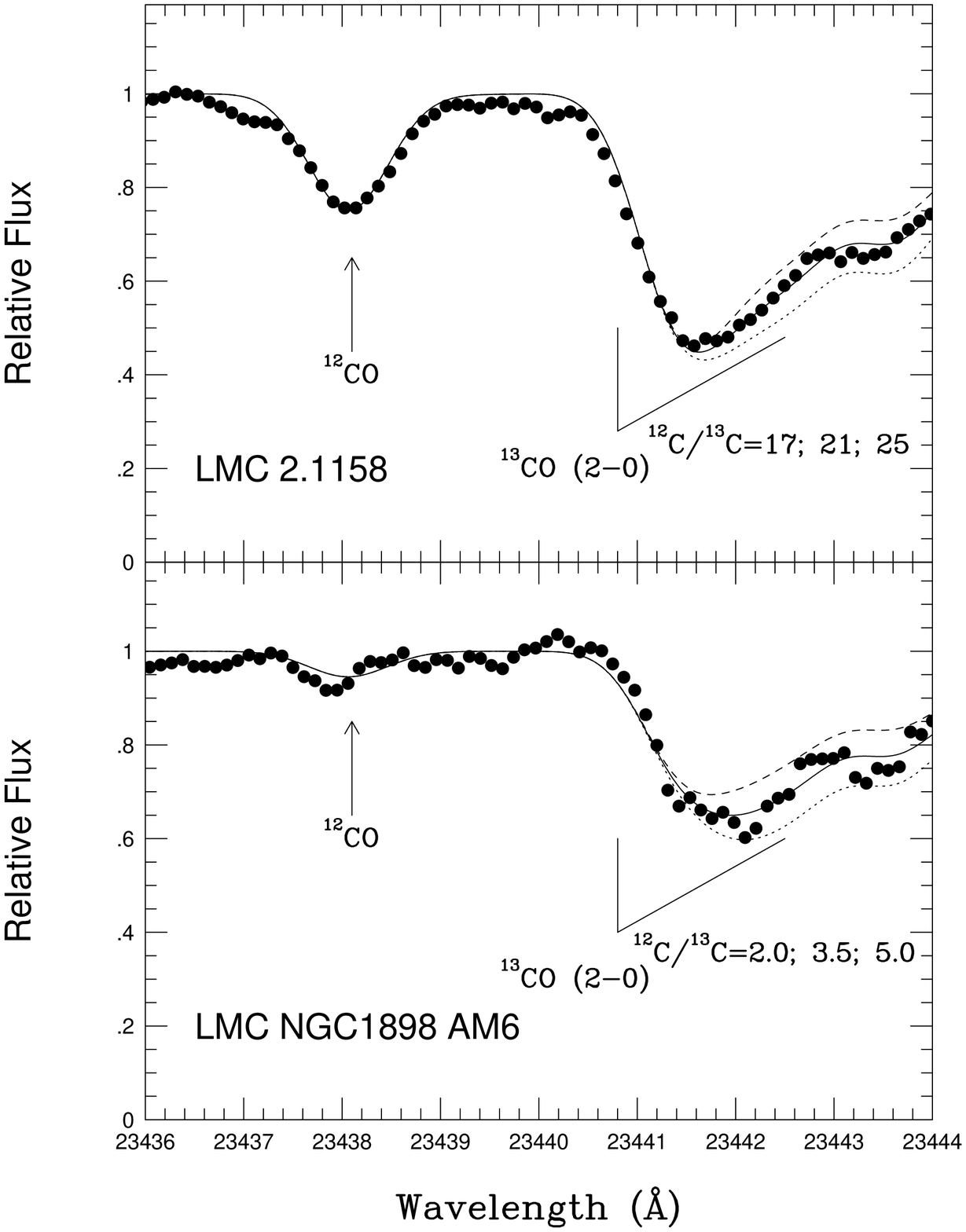]{Examples of synthetic spectral fits to the blended
$^{13}$CO (2-0) bandhead in two LMC red giants.  The top panel shows
a field star (2.1158) that has an estimated mass of 3M$_{\odot}$, with
a metallicity of [Fe/H]= -0.34, and having a fairly large $^{12}$C/$^{13}$C
ratio, typical of the slightly higher mass red giants in this sample.
The bottom panel illustrates the $^{13}$CO bandhead in the most metal-poor
star in the sample, the cluster member AM6 in NGC1898, with [Fe/H]= -1.13.
This is an old LMC red giant, with an estimated mass of 1M$_{\odot}$,
and having a low isotopic ratio ($^{12}$C/$^{13}$C= 3.5); the trend
in the LMC of lower carbon isotope ratios with lower mass and metallicity
is in the same sense as in the Milky Way. 
\label{fig6}}

\figcaption[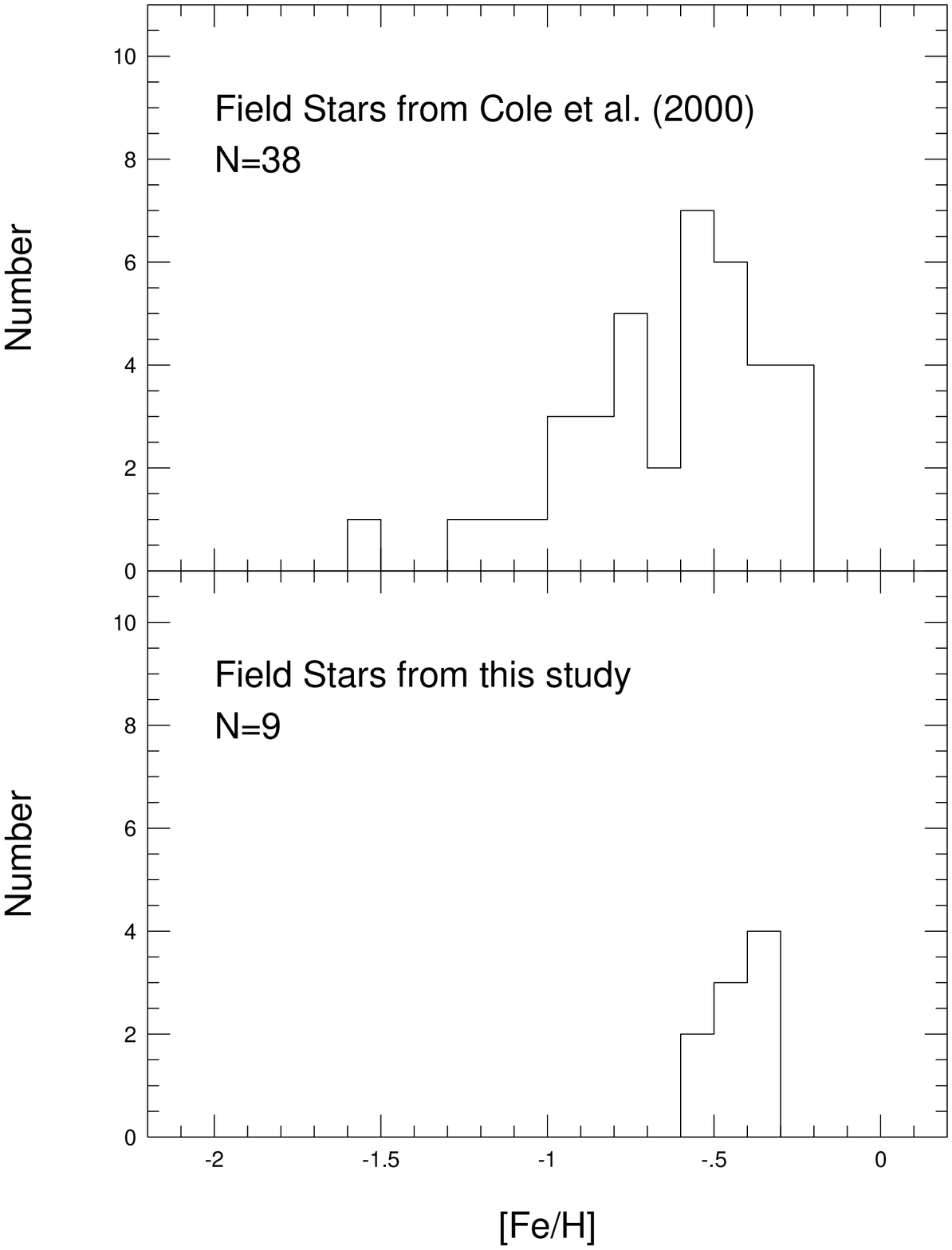]{A comparison of field star iron abundances from the
large study by Cole et al. (2000- top panel) and for the field LMC red
giants analyzed here.  The Cole et al. results are derived from 
low-resolution spectra of the Ca II IR triplet, while [Fe/H] for the
stars here are from three Fe I lines in the H-band.  Our current
sample of 9 field stars is too small to compare to the shape of the
metallicity distribution as defined by the Cole et al. (2000) sample;
however, the average [Fe/H] of our sample agrees very well with that
expected from the Cole et al. distribution. 
\label{fig7}}

\figcaption[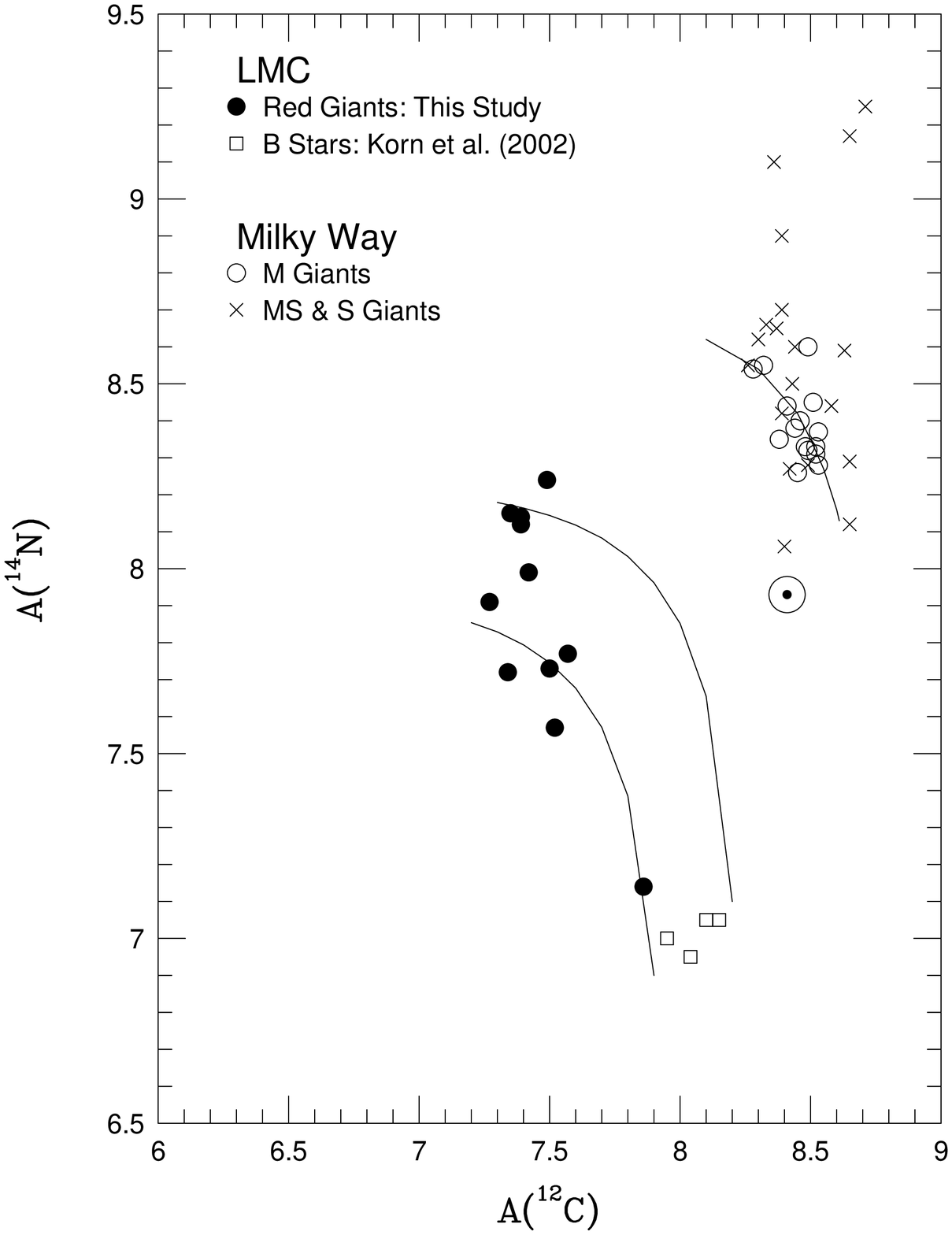]{Abundances of $^{14}$N versus $^{12}$C (on the scale
of A(x)= log[N(x)/N(H)] + 12) for the LMC red giants along with other
LMC samples, as well as Galactic red giants (from Smith \& Lambert 1985;
1986; 1990).  The solid curves illustrate 
``CN mixing lines'': since the CN cycle, operating by itself,
will conserve the total 
C plus N nuclei, the curves represent these relations.  No account
is taken of some carbon cycled into $^{13}$C, however in the extreme case
of the steady state CN-cycle, this would shift the $^{12}$C abundances
by, at most, 0.1 dex towards lower values.  In the case of the Galactic
M-giants, they are approximated reasonably well by CN-cycle mixing in
near-solar metallicity material.  Some of the very $^{14}$N-rich MS or S
stars have converted some $^{16}$O, or $^{12}$C produced during 
$^{4}$He-burning, into nitrogen.  The LMC red giants fall along CN mixing
lines that are defined well by initial carbon and nitrogen abundances
as derived by Korn et al. (2002) for main-sequence B-stars
in the young LMC cluster NGC2004.  
\label{fig8}}

\figcaption[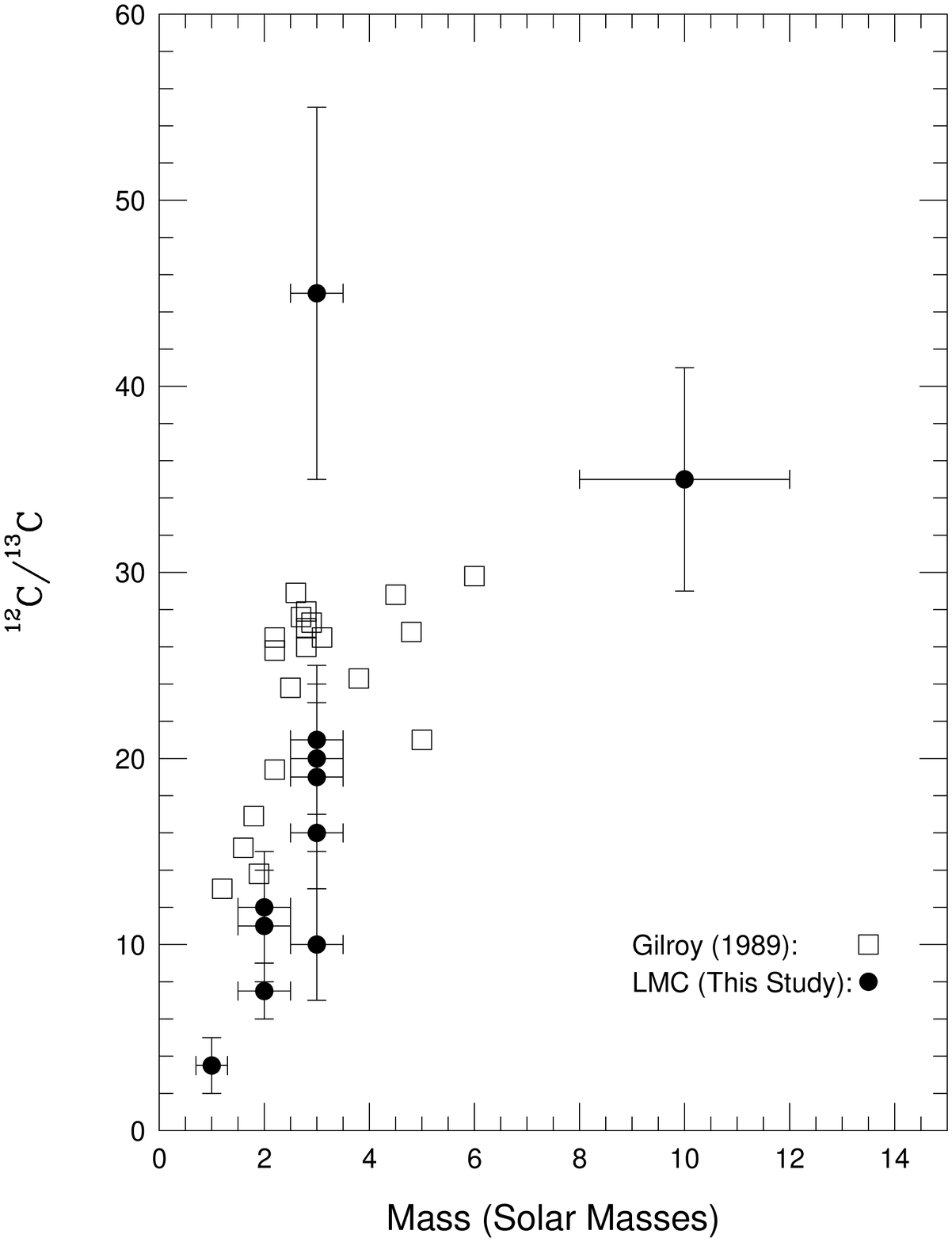]{Carbon-12 to carbon-13 ratios versus stellar
mass in the LMC red giants and in Galactic open-cluster giants from
Gilroy (1989).  The LMC masses are estimated from the evolutionary tracks
of Schaerer et al. (1993).  The trend of a lower $^{12}$C/$^{13}$C
ratio with a lower stellar mass observed in the LMC sample of giants
is in qualitative agreement with the result found by Gilroy (1989)
for the Galactic giants.  The noticable offset between LMC and Galactic
trends may be due to an increase in extra-mixing efficiency on the
giant branch with decreasing metallicity as argued by Charbonnel et
al. (1998).    
\label{fig9}}

\figcaption[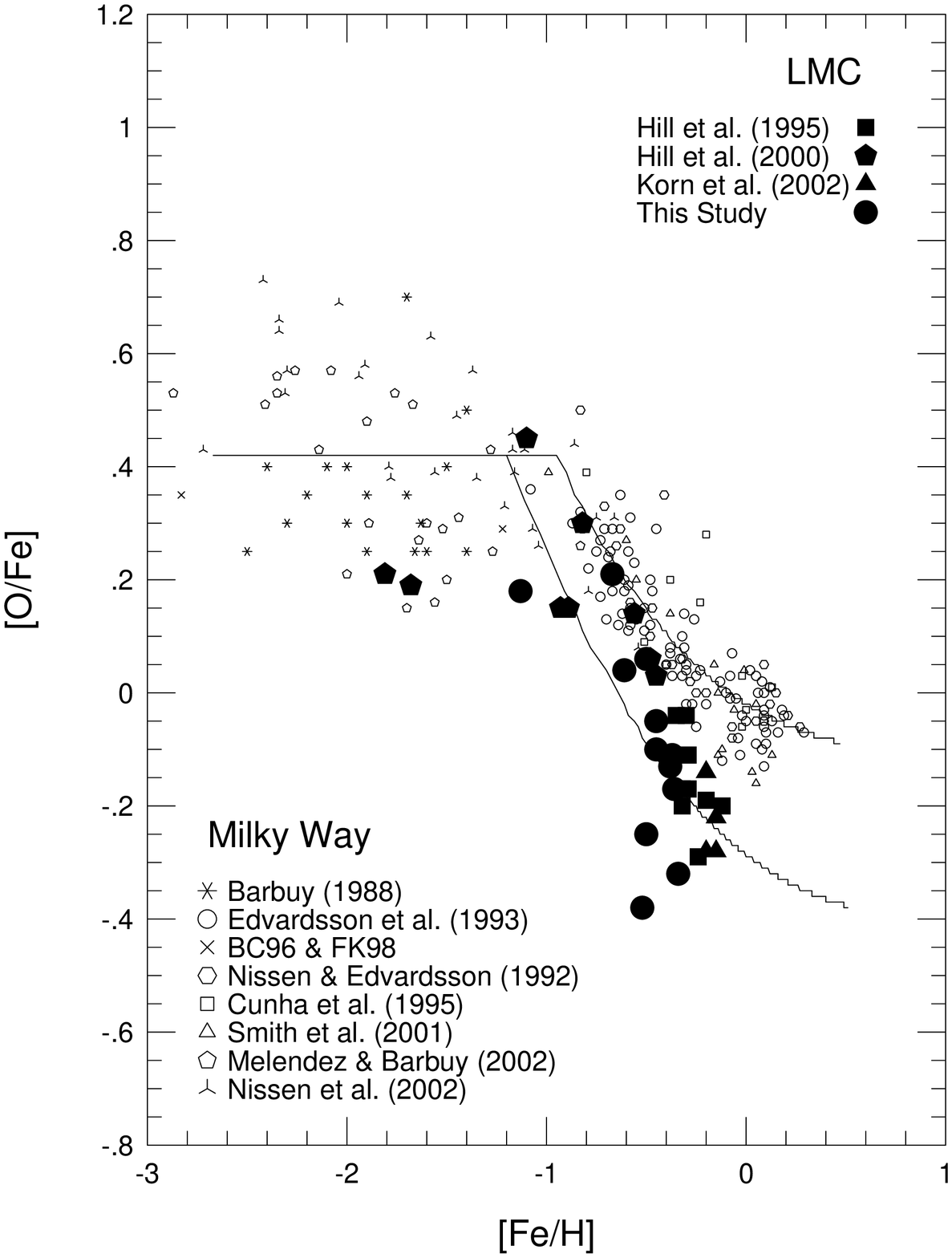]{Values of [O/Fe] versus [Fe/H] for the LMC red
giants, along with other samples of LMC and Galactic stars.  Most of
the Galactic results rely on the [O I] line near 6300\AA,
weak O I lines near 6158\AA, or 
IR vibration-rotation OH lines (as used here).  The LMC F-supergiants
from Hill et al. (1995) rely on the weak O I lines near 6158\AA,
while the Korn et al. (2002) results for the hot B-stars use lines
of O II and Fe III.  All three samples of LMC stars indicate a trend
of [O/Fe] that falls $\sim$ 0.2 dex below the Galactic trend at
[Fe/H]$\sim$ -0.5 to 0.0.  The fact that three quite different samples
of stars provide such a consistent result is a strong indication
that [O/Fe] is low in the LMC compared to the Mikly Way at near-solar
values of [Fe/H].  The continuous curves are simple models of chemical
evolution in which supernovae of type II and Ia add O and Fe into
mixed gas.  The LMC curve differs from the Milky Way curve by having
a SN II rate that is 3 times lower and a SN Ia rate that is 2 times lower. 
\label{fig10}}


\clearpage

\begin{thebibliography}{}

\bibitem[]{} Aaronson, M., \& Mould, J. 1985, ApJ, 288, 551

\bibitem[]{} Allende Prieto, C., Lambert, D. L., \& Asplund, M. 2002, ApJ,
573, L137 

\bibitem[]{} Arnett, D. 1999, Supernovae and Nucleosynthesis (Princeton:
Princeton University Press)

\bibitem[]{} Asplund, M., Gustafsson, B., Kiselman, D., \& Eriksson, K.
1997, A\&A, 318, 521

\bibitem[]{} Asplund, M., \& Garcia P\'erez, A. E. 2001, A\&A, 372, 601 

\bibitem[]{} Balachandran, S. C. \& Carney, B. W. 1996, AJ 111, 946  

\bibitem[]{} Barbuy, B. 1988, A\&A, 191, 121

\bibitem[]{} Barbuy, B., de Freitas Pacheco, J. A., \& Castro, S. 1994,
A\&A, 283, 32 

\bibitem[]{} Bessell, M.S., \& Brett, J.M. 1988, PASP, 100, 1134

\bibitem[]{} Bessell, M.S., Castelli, F., \& Plez, B. 1998, A\&A, 333, 231

\bibitem[]{} Bica, E., Geisler, D., Dottori, H., Clari\'a, J.J., Piatti, A.E.,
\& Santos, J.F.C. Jr. 1998, AJ, 116, 723

\bibitem[]{} Carpenter, J.M. 2001, AJ, 121, 2851

\bibitem[]{} Charbonnel, C. 1994, A\&A, 282, 811

\bibitem[]{} Charbonnel, C., Brown, J. A., \& Wallerstein, G. 1998,
A\&A, 332, 204

\bibitem[]{} Cole, A. A., Smecker-Hane, T. A., \& Gallagher, J. S., III
2000, AJ, 120, 1808

\bibitem[]{} Costes, M., Naulin, C., \& Dorthe, G. 1990, A\&A, 232, 270

\bibitem[]{} Cunha, K., Smith, V.V., \& Lambert, D.L. 1995, ApJ, 452, 634 

\bibitem[]{} Da Costa, G. S. 1988 in Harlow Shapley Symposium on Globular
Cluster Systems in Galaxies, edited by J. E. Grinlay and A. G. Davis 
Phillip (Reidel, Dordrecht), p. 191 

\bibitem[]{} D'Antona, F., \& Mazzitelli, I. 1996, ApJ, 470, 1093

\bibitem[]{} Dopita, M.A., Vassiliadis, E., Meatheringham, S.J., Bohlin, K;C;,
Ford, H.C., Harrington, J.P., Wood, P.R., Stecher, T.P., \& Maran, S.P.
1996, ApJ, 460, 320 

\bibitem[]{} Edvardsson, B., Andersen, J., Gustafsson, B.,
Lambert, D.L., Nissen, P.E. \&  Tomkin, J. 1993, A\&A, 275, 101

\bibitem[]{} Geisler, D., Bica, E., Dottori, H., Clari\'a, J.J., Piatti, A.E.,
\& Santos, J.F.C. Jr. 1997, AJ, 114, 1920

\bibitem[]{} Gilroy, K. K. 1989, ApJ, 347, 835 

\bibitem[]{}  Goldman, A., Shoenfeld, W. G., Goorvitch, D., Chackerian C. Jr.,
Dothe, H., M\'elen, F., Abrams, M. C. \&  Selby, J. E. A. 1998, JQSRT, 59, 45

\bibitem[]{} Goorvitch, D. 1994, ApJS, 95, 535

\bibitem[]{} Gratton, R.G., \& Sneden, C. 1991, A\&A, 241, 501

\bibitem[]{} Grebel, E. K., Roberts, W. J. and Van de Rydt, F. 1994, 
Third ESO/CTIO 
Workshop, La Serena, Chile

\bibitem[]{} Gustafsson, B., Bell, R. A., Eriksson, K., \& Nordlund, A.
1975, A\&A, 42, 407

\bibitem[]{} Hill, V., Andrievsky, S., \& Spite, M. 1995, A\&A, 293, 
347

\bibitem[]{} Hill, V., Francois, P., Spite, M., Primas, F., \& Spite, F.
2000, A\&A, 364, L19

\bibitem[]{} Hill, V., Plez, B., Cayrel, R., Beers, T. C., Nordstrom, B.,
Andersen, J., Spite, M., Spite, F., Barbuy, B., Bonifacio, P., Depagne, E.,
Francois, P., Primas, F. 2002, A\&A, 387, 560

\bibitem[]{} Hinkle, K. H., Wallace, L., \& Livingston, W. 1995,
Infrared Atlas of the Arcturus Spectrum, 0.9 - 5.3 $\mu$m (San Francisco:
Astronomical Society of the Pacific)

\bibitem[]{} Hinkle, K. H., Cuberly, R., Gaughan, N., Heynssens, J.,
Joyce, R., Ridgway, S., Schmitt, P., \& Simmons, J. E. 1998,
Proc. SPIE, 3354, 810

\bibitem[]{} Hinkle, K. H., Joyce, R. R., Sharp, N., \& Valenti,
J. A. 2000, Proc. SPIE, 4008, 720

\bibitem[]{} Hinkle, K. H., Blum, R., Joyce, R. R., Ridgway, S. T.,
Rodgers, B., Sharp, N., Smith, V., Valenti, J., \& van der
Bliek, N. 2002, Proc. SPIE, 4834, in press 

\bibitem[]{} Holweger, H. 2001, in Solar and Galactic Composition
(New York: American Institute of Physics), ed. R. F. Wimmer-Schweingruber,
p.23 

\bibitem[]{} Huber, K. P. \& Herzberg, G. 1979, Constants of Diatomic 
Molecules, Van Nostrand Reinhold, New York.

\bibitem[]{} Iben, I., Jr. 1964, ApJ, 140, 1631 

\bibitem[]{} Keller, L. D., Pilachowski, C. A., \& Sneden, C. 2001,
AJ, 122, 2554

\bibitem[]{} Korn, A. J., Becker, S. R., Gummersbach, C. A., \& Wolf, B.
2000, A\&A, 353, 655

\bibitem[]{} Korn, A. J., Keller, S. C., Kaufer, A., Langer, N., Przybilla,
N., Stahl, O., \& Wolf, B. 2002, A\&A, 385, 143

\bibitem[]{} Kurucz, R.L., \& Bell, B. 1995, CD-ROM 23

\bibitem[]{} Luck, R. E., Moffett, T. J., Barnes, T. G., III, \& Gieren,
W. P. 1998, AJ, 115, 605  

\bibitem[]{} McWilliam, A. 1990, ApJS, 74, 1075

\bibitem[]{} McWilliam, A. 1997 ARA\&A 35, 503

\bibitem[]{} Mel\'endez, J., Barbuy, B., \& Spite, F. 2001, ApJ, 556, 858

\bibitem[]{} Mel\'endez, J., \& Barbuy, B. 2002, ApJ, 575, 474 

\bibitem[]{} Nissen, P.E., \& Edvardsson, E. 1992, A\&A, 261, 255

\bibitem[]{} Nissen, P.E., Primas, F., Asplund, M., \& Lambert, D.L. 2002,
A\&A, 390, 235

\bibitem[]{} Olszewski, E.W., Schommer, R.A., Suntzeff, N.B., 
\& Harris, H.C. 1991, AJ, 101, 515 

\bibitem[]{} Pagel, B. E. J. \& Tautvai\v{s}ien\.{e}, G. 1995, 
MNRAS, 276, 505

\bibitem[]{} Plez, B., Brett, J.M., \& Nordlund, {\rm \AA}. 1992, A\&A, 256,
551

\bibitem[]{} Prevot, L., Andersen, J., Ardeberg, A., Benz, W.,
Imbert, M., Lindgren, H., Martin, N., Maurice, E., Mayor, M.,
Nordstrom, B., Rebeirot, E., \& Rosseau, J. 1985, A\&AS, 62, 23

\bibitem[]{} Russell, S. C., \& Dopita, M. A. 1992, ApJ, 384, 508

\bibitem[]{} Schaerer, D., Charbonnel, C., Meynet, G., Maeder, M., \&
Schaller, G. 1993,
         A\&AS, 102, 339 

\bibitem[]{} Shetrone, M.D., C\^ot\'e, P., \& Sargent, W.L.W. 2001, ApJ, 548,
592

\bibitem[]{} Smith, V.V., \& Lambert, D.L. 1985, ApJ, 294, 326 

\bibitem[]{} Smith, V.V., \& Lambert, D.L. 1986, ApJ, 311, 843 

\bibitem[]{} Smith, V.V., \& Lambert, D.L. 1990, ApJS, 72, 387 

\bibitem[]{} Smith, V.V., Suntzeff, N.B., Cunha, K., Gallino, R., Busso, M.,
Lambert, D.L.,
\& Straniero, O.  2000, AJ, 119, 1239 

\bibitem[]{} Smith, V.V., Cunha, K., \& King, J. 2001, AJ, 122, 370 

\bibitem[]{} Sneden, C. 1973, ApJ, 184, 839

\bibitem[]{} Spite, M., Hill, V., Primas, F., Francois, P,. \& Spite, F.
2001, New Astronomy Reviews, 45, 557

\end{thebibliography}
\end{document}